\documentclass{ws-rv9x6}

\begin{document}

\setcounter{chapter}{0}

\chapter{QUANTUM PHASE TRANSITIONS\\
         IN ALTERNATING TRANSVERSE ISING CHAINS}

\markboth{O. Derzhko}{Quantum Phase Transitions
                      in Alternating Transverse Ising Chains}

\author{Oleg Derzhko}

\address{Institute for Condensed Matter Physics\\
         of the National Academy of Sciences of Ukraine\\
         1 Svientsitskii Street, L'viv-11, 79011, Ukraine\\
         E-mail: derzhko@icmp.lviv.ua}

\begin{abstract}
This chapter is devoted to a discussion
of quantum phase transitions
in regularly alternating spin-$\frac{1}{2}$ Ising chain in a transverse field.
After recalling
some generally-known topics
of the classical (temperature-driven) phase transition theory
and
some basic concepts
of the quantum phase transition theory
I pass to the statistical mechanics calculations
for a one-dimensional spin-$\frac{1}{2}$ Ising model in a transverse field,
which is the simplest possible system
exhibiting the continuous quantum phase transition.
The essential tool for these calculations
is the Jordan-Wigner fermionization.
The latter technique
being completed by the continued fraction approach
permits to obtain analytically
the thermodynamic quantities
for a ``slightly complicated'' model
in which the intersite exchange interactions and on-site fields
vary regularly along a chain.
Rigorous analytical results
for the ground-state and thermodynamic quantities,
as well as
exact numerical data for the spin correlations
computed for long chains
(up to a few thousand sites)
demonstrate how the regularly alternating bonds/fields
effect the quantum phase transition.
I discuss in detail the case of period 2,
swiftly sketch the case of period 3
and
finally summarize
emphasizing the effects
of periodically modulated Hamiltonian parameters
on quantum phase transitions
in the transverse Ising chain
and in some related models.
\end{abstract}

\section{Classical and Quantum Phase Transitions
         \label{secd1}}

In this chapter
I consider the effects of regular alternation of the Hamiltonian parameters
on the quantum phase transition
inherent in a one-dimensional spin-$\frac{1}{2}$ Ising model in a transverse field.
Before starting to discuss this issue
let me recall some common wisdoms from statistical physics.
One of the aims of statistical physics
is to describe different phases
which may occur in many-particle systems
and the transitions between different phases,
i.e. the phase transitions.
We often face phase transitions in everyday life.
Melting of ice,
boiling of water
or vanishing of the magnetic properties of iron after heating
are well-known phenomena for everyone.
Normally
we associate the changes in properties of a substance
as the temperature varies.
However,
such changes may occur at a fixed finite temperature
while some other parameter
(such as presssure)
varies.

P.~Ehrenfest proposed a classification of phase transitions.
In the Ehrenfest classification of phase transitions
we say that the phase transition is of the order one
if the free energy
is continuous across the phase transition
whereas its first derivatives with respect to temperature and other variables
(for example, pressure)
are discontinuous.
Similarly,
we say that the phase transition is of the order two
(three)
if the free energy and its first derivatives
(first two derivatives)
are continuous across the phase transition
whereas the second
(third)
derivatives
are discontinuous.
A few years later
L.~Landau proposed
to consider discontinuous or continuous phase transitions.
In the Landau sense
the discontinuous (continuous) phase transition
is characterized by a discontinuous (continuous) change in the order parameter
and, therefore,
it is usually viewed as of the first- (second- or higher-) order.
In the first-order phase transitions
the two phases coexist
at the phase transition temperature.
Thus, if ice is melting one observes two phases,
i.e., liquid and solid,
which coexist
at the temperature of phase transition.
This is an example of discontinuous phase transition.
In the second-order (and higher-order) phase transitions
the two phases do not coexist.
An example is the Curie point of a ferromagnet
above which the magnetic moment of a material vanishes.
Below the Curie temperature $T_c$
we can observe only the ferromagnetic phase
which continuosly disappears at $T_c$.
Above $T_c$
we can observe only the paramagnetic phase.
Some phase transitions are not in accord
with the naively applied classification rules.
Thus,
the Bose-Einstein condensation
in an ideal Bose gas
is accompanied by a kink
in the temperature dependence of specific heat
at the Bose-Einstein condensation temperature
but is viewed as a first-order phase transition.\cite{001}
In what follows
our focus will be on the continuous phase transitions
which were studied very intensively
in the last century,
especially in the last four decades.

Statistical mechanics
provides some microscopic models
of the continuous phase transitions.
One of such models
\index{Ising model}
was invented by E.~Ising\cite{002}
about 80 years ago.
It became especially well-known
after L.~Onsager\cite{003} had found
an exact solution of the model
in two dimensions.
Let me introduce the Ising model
and fix the notations.
The model consists of magnetic moments or spins
which may have two values $-\frac{1}{2}$ and $\frac{1}{2}$
and
which interact with the nearest neighbors.
The Hamiltonian of the model on a square lattice
of $N_xN_y=N$ sites
can be written in the form
\begin{eqnarray}
H
=-\sum_{i_x=1}^{N_x}\sum_{i_y=1}^{N_y}
\left(
J_hs_{i_x,i_y}^zs_{i_x+1,i_y}^z
+J_vs_{i_x,i_y}^zs_{i_x,i_y+1}^z
\right)
\label{1.01}
\end{eqnarray}
where
$s_{i_x,i_y}^z$
is the spin variable attached to the site
$i_x,i_y$,
and $J_h$ and $J_v$ are the exchange interactions
in horizontal and vertical directions,
respectively.
If the exchange interaction in (\ref{1.01}) is positive,
the same value of spin variables at neighboring sites
is favorable,
i.e. the exchange interaction is ferromagnetic.
The spin variable $s^z$
may be presented by a half of the Pauli matrix
$$
\left(
\begin{array}{cc}
1 &  0 \\
0 & -1
\end{array}
\right)
$$
and the canonical partition function
which determines the thermodynamics of the model
reads
\begin{eqnarray}
Z
=\sum_{\left\{s^z_{i_x,i_y}=\pm \frac{1}{2}\right\}}
\exp\left(-\beta H\right)
={\mbox{Tr}}\exp\left(-\beta H\right),
\label{1.02}
\end{eqnarray}
where $\beta=\frac{1}{kT}$
is the inverse temperature.
We are also interested in the spin correlation functions
$
\langle s^z_{i_x,i_y}s^z_{j_x,j_y}\rangle,
$
where the canonical average means
\begin{eqnarray}
\langle\left(\ldots\right)\rangle
=\frac{1}{Z}{\mbox{Tr}}\left(\exp\left(-\beta H\right)\left(\ldots\right)\right).
\nonumber
\end{eqnarray}
The spin correlation functions
in the limit of infinitely large intersite distances
yield the magnetization per site
\begin{eqnarray}
m^z
=\frac{1}{N_xN_y}
\sum_{i_x=1}^{N_x}\sum_{i_y=1}^{N_y}
\langle s^z_{i_x,i_y}\rangle
\label{1.03}
\end{eqnarray}
which plays the role of the order parameter.
Due to the seminal study of L.~Onsager
we understand in great detail
the properties of the two-dimensional Ising model.

At zero temperature $T=0$ all spins have the same value,
say,
$\frac{1}{2}$,
and $m^z=\frac{1}{2}$.
As the temperature becomes nonzero
some of spin variables
due to temperature fluctuations
have the value $-\frac{1}{2}$,
and thus $m^z$ becomes smaller than $\frac{1}{2}$.
Quantitatively the temperature fluctuations of magnetic moment
can be characterized by
$$\left\langle
\left(
\sum_{i_x=1}^{N_x}\sum_{i_y=1}^{N_y}
\left(s^z_{i_x,i_y}-m^z\right)
\right)^2
\right\rangle.$$
At the critical temperature $T_c$
the temperature fluctuations completely destroy the order:
the average numbers of spins
having the values $\frac{1}{2}$ and $-\frac{1}{2}$
are the same
and $m^z=0$.
Above $T_c$
the average numbers of spins
having the values $\frac{1}{2}$ and $-\frac{1}{2}$
remain the same
and $m^z=0$.
Analytical calculations predict
the logarithmic singularity of the specific heat
in the vicinity of $T_c$
\begin{eqnarray}
c\sim\ln\left\vert T-T_c\right\vert,
\label{1.04}
\end{eqnarray}
and
the vanishing of the order parameter $m^z$
while $T_c-T\to +0$
as
\begin{eqnarray}
m^z\sim\left( T_c-T\right)^{\frac{1}{8}}.
\label{1.05}
\end{eqnarray}
The correlation length $\xi$
which characterizes the long-distance behavior
of spin correlations,
\begin{eqnarray}
\langle s^z_{{\bf{i}}} s^z_{{\bf{j}}} \rangle
\sim \exp\left(-\frac{\left\vert{\bf{i}}-{\bf{j}}\right\vert}{\xi}\right),
\;\;\;
\left\vert{\bf{i}}-{\bf{j}}\right\vert\to\infty,
\label{1.06}
\end{eqnarray}
diverges in the vicinity of $T_c$ as
\begin{eqnarray}
\xi\sim\left\vert T-T_c \right\vert^{-1}.
\label{1.07}
\end{eqnarray}

In one dimension $T_c=0$.
At any infinitesimally small temperature
the temperature fluctuations destroy
the long-range order
which exists only at zero temperature.
The exact solution\cite{003_4}
for the two-point spin correlation functions reads
\begin{eqnarray}
\langle s_j^zs_{j+n}^z\rangle
=\frac{1}{4}
\left(\tanh\frac{\beta J}{4}\right)^n
= \frac{1}{4}
\exp\left(-\frac{n}{\xi}\right),
\nonumber\\
\frac{1}{\xi}=\ln\coth\frac{\beta J}{4},
\label{1.08}
\end{eqnarray}
and hence
the correlation length $\xi$ diverges
while $T\to T_c=0$
as
\begin{eqnarray}
\xi\sim \exp\frac{J}{2kT}.
\label{1.09}
\end{eqnarray}

We do not know the exact solution
of the Ising model in three dimensions
although qualitatevilely
the briefly sketched picture for a two-dimensional case
remains valid.

Numerous experimental studies
of different substances
in the vicinity of the continuous phase transition points
show
that the critical behavior is characterized by a set of exponents
which may be identical for different substances.
This remarkable result of universality
urged the researchers to proceed
elaborating scaling concepts,
establishing scaling relations
and developing renormalization ideas
and many systematic renormalization-group schemes
in order to calculate critical exponents.
These issues constitute a basic course
in the theory of classical
(i.e. temperature-driven)
continuous phase transitions.\cite{004,005}

\begin{figure}[th]
\centerline{\psfig{file=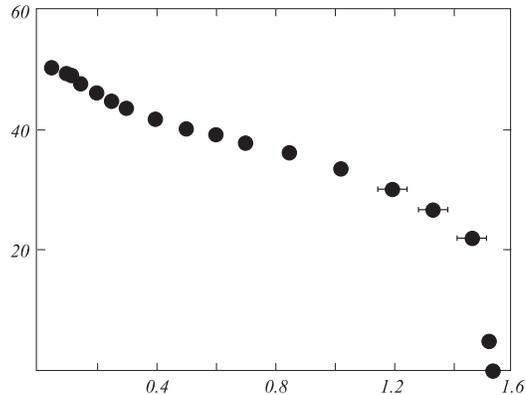,width=2.7in,angle=0}}
\vspace*{8pt}
\caption
{The temperature -- transverse field phase diagram of LiHoF$_4$.
The temperature $T$ (the horizontal axis)
is measured in K,
the (transverse) magnetic field $H^t$ (the vertical axis)
is measured in kOe.
Experimental data
for the ferromagnetic transition
(filled circles)
are obtained via magnetic susceptibility measurements.}
\label{fig01}
\end{figure}

In what follows
I shall not speak about temperature-driven continuous phase transitions.
I wish to focus on
\index{quantum phase transition}
the quantum continuous phase transitions.
Let me start from a brief discussion of the experiment
performed by D.~Bitko, T.~F.~Rosenbaum and G.~Aeppli\cite{006}
which demonstrates
how a phase transition
may be driven by entirely quantum rather than temperature fluctuations.
D.~Bitko et al carried out their measurements
for a model magnet
lithium holmium fluoride LiHoF$_4$
in the external field $H^t$,
which turns out to be the experimental realization
of the Ising magnet in a transverse magnetic field.
At low temperatures
(below 2~K),
the magnetic properties arise
owing to the magnetic dipolar interaction of the spins
of neighboring holmium ions Ho$^{3+}$.
These spins prefer to be directed
either up or down
with respect to a certain crystalline axis
and present the three-dimensional ferromagnetic Ising model.
D.~Bitko et al
investigated the behavior of
LiHoF$_4$
as a function of temperature $T$
and an external magnetic field $H^t$
applied normally to the Ising axis.
For this purpose
they measured the real and the imaginary parts of the magnetic susceptibility
along the Ising axis
at different temperatures establishing $T_c$ or $H^t_c$.
Their findings are summarized in Fig.~\ref{fig01}.

Let us discuss the experimentally measured phase diagram
(Fig.~\ref{fig01})
of the system
which is described by the Hamiltonian
of the Ising magnet in a transverse magnetic field
\index{Ising model in a transverse field}
\begin{eqnarray}
H
=\sum_{i,j}J_{ij}s^z_is^z_j
-\Omega\sum_i s^x_i,
\label{1.10}
\end{eqnarray}
where
$s^\alpha$, $\alpha=x,y,z$
are halves of the Pauli matrices
\begin{eqnarray}
\sigma^x
=
\left(
\begin{array}{cc}
0 & 1 \\
1 & 0
\end{array}
\right),
\;\;\;
\sigma^y
=
\left(
\begin{array}{cc}
0          & -{\mbox{i}} \\
{\mbox{i}} & 0
\end{array}
\right),
\;\;\;
\sigma^z
=
\left(
\begin{array}{cc}
1 &  0 \\
0 & -1
\end{array}
\right),
\label{1.11}
\end{eqnarray}
$J_{ij}$s are the Ising exchange couplings
and
$\Omega$ is a transverse field.
Consider the behavior of the system as temperature increases
at $H^t=0$
(the horizontal axis in Fig.~\ref{fig01}).
At $T=0$,
all spins are pointed,
say,  up
(i.e., fully polarized ferromagnetic state).
As the temperature becomes nonzero,
the temperature fluctuations
cause some spins of the ground-state configuration to flip,
i.e. some spins become pointed down.
As the temperature increases,
the number of flipped spins increases
and, at $T_c=1.53$~K,
the numbers of up and down spins become equal.
The numbers of spin up and spin down
are the same for all temperatures above $T_c$.
This is a scenario
of the conventional temperature-driven continuous phase transition
from the ferromagnetic to the paramagnetic phase
with the Curie temperature $T_c=1.53$~K.

Consider further
what happens
while we are moving along the vertical axis in Fig.~\ref{fig01},
i.e. while we increase the transverse field $H^t$
at zero temperature $T=0$.
We immediately observe
that the existing order at zero temperature
may be destroyed in a completely different manner
even at $T=0$.
Applying transverse field $H^t$
we allow tunneling between spin-down and spin-up states.
And if
$H^t>H^t_c$,
the ground state becomes paramagnetic even at $T=0$.
Hence,
the quantum fluctuations destroy the order.
In other words,
the phase transition between ferromagnetic and paramagnetic phases
described by the order parameter
$\frac{1}{N}\sum_i\langle s_i^z\rangle$
is driven entirely by quantum fluctuations.
For both $T$ and $H^t$ nonzero,
D.~Bitko et al
found experimentally a line of continuous phase transitions
which separates the ferromagnetic phase
(lower region in Fig.~\ref{fig01})
from the paramagnetic phase
(upper region in Fig.~\ref{fig01}).
For example,
at $T=0.100$~K
they found
$H^t_c=49.3$~kOe.
Thus,
the classical phase transition at $H^t=0$
and the quantum phase transition at $T=0$
are connected by a line of continuous phase transitions.
To develop a correct theoretical description
of the equally important quantum and temperature fluctuations in this region
has been a focus of many recent studies.

For further discussions on this issue
as well as for other examples of quantum phase transitions
see
the review article for a general science audience\cite{007}
and
the book\cite{008} of S.~Sachdev.

\section{Spin-$\frac{1}{2}$ Ising Chain in a Transverse Field
         as the Simplest Model for the Quantum Phase Transition Theory
         \label{secd2}}

In early sixties of the last century
E.~Lieb, T.~Schultz and D.~Mattis\cite{009}
(see also the paper by S.~Katsura\cite{010})
suggested a new exactly solvable model,
the so-called,
\index{spin-$\frac{1}{2}$ $XY$ chain}
spin-$\frac{1}{2}$ $XY$ chain.
In particular case
it transforms
\index{Ising chain in a transverse field}
into the one-dimensional spin-$\frac{1}{2}$ Ising model
in a transverse field;
this case was examined in detail
several years later by P.~Pfeuty.\cite{011}
The long-known results for the Ising chain in a transverse field
may be viewed in the context of the quantum phase transition theory.
The transverse Ising chain
is apparently the simplest model
exhibiting the continuous quantum phase transition
which can be studied in much detail
since many statistical mechanics quantities for that model
are amenable for rigorous calculations.

I begin to discuss these results introducing the model.
It consists of $N\to\infty$ spins $\frac{1}{2}$
(which are represented by halves of the Pauli matrices)
which are arranged in a row.
Only the neighboring spins interact
via the Ising exchange interaction.
Moreover,
the spins interact with an external transverse field.
The Hamiltonian of the model may be written as follows:
\begin{eqnarray}
H=\sum_{n=1}^N\Omega s_n^z
+\sum_{n=1}^N Js_n^xs_{n+1}^x.
\label{2.01}
\end{eqnarray}
We may impose periodic (cyclic) boundary conditions
assuming
$s^\alpha_{N+1}=s^\alpha_1$
or open (free) boundary conditions
assuming
the last term in the second sum in (\ref{2.01})
to be zero.
The Ising chain without transverse field,
$\Omega=0$,
exhibits the long-range order at zero temperature $T=0$
with the order parameter
$\langle s^x\rangle
=\frac{1}{N}\sum_i\langle s_i^x\rangle$
which equals,
say,
$\frac{1}{2}$
(we assume the exchange interaction in (\ref{2.01}) to be ferromagnetic,
$J<0$).
The long-range order is immediately destroyed,
i.e. $\langle s^x\rangle =0$,
for any nonzero temperature $T>0$. However,
the Ising magnetization
$\langle s^x\rangle$
can be destroyed at $T=0$
by quantum fluctuations,
when we switch on in (\ref{2.01})
the transverse field $\Omega\ne 0$.
Small values of $\Omega$ reduce
the transverse magnetizations $\langle s^x\rangle$
and after $\Omega$ exceeds the critical value
$\Omega_c=\frac{\vert J\vert}{2}$
the Ising magnetization becomes zero.
That is the scenario which was discussed above
in connection with the low-temperature properties of LiHoF$_4$.
The advantage of the introduced model (\ref{2.01})
is a possibility to follow in great detail
the quantum phase transition tuned by $\Omega$.

\begin{figure}[th]
\centerline{\psfig{file=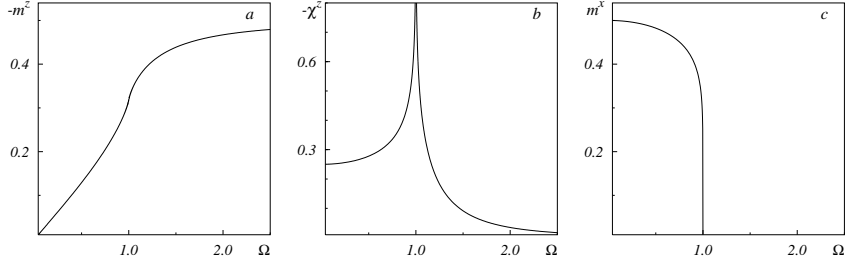,width=4.6in,angle=0}}
\vspace*{8pt}
\caption{The ground-state dependences
of the transverse magnetization $m^z$ (a),
the static transverse susceptibility $\chi^z$ (b),
and the longitudinal magnetization $m^x$ (c)
on the transverse field $\Omega$
for the transverse Ising chain (\ref{2.01})
($J=-2$).}
\label{fig02}
\end{figure}

Just after introducing the spin-$\frac{1}{2}$ $XY$ chains it was recognized
that there was an intimate connection
between  the transverse Ising chain
and the square-lattice Ising model
(see, for example, a redirevation of the Onsager solution
using the Jordan-Wigner fermionization\cite{012}).
The relationship between these models
was demonstrated explicitly by M.~Suzuki.\cite{013,014}
Consider the square-lattice Ising model (\ref{1.01})
with
the strength of horizontal interactions $J_h>0$
and
with
the strength of vertical interactions $J_v>0$
and the transverse Ising chain (\ref{2.01})
with the exchange interaction $J<0$
and
with the transverse field $\Omega$.
The thermodynamic properties
of the square-lattice Ising model
are
equivalent to the ground-state properties
of the transverse Ising chain
under the relations\cite{013,014,015}
\begin{eqnarray}
J_h\to 0,
\;\;\;
J_v\to\infty,
\;\;\;
\frac{\exp\left(-\frac{J_v}{2kT}\right)}{\frac{J_h}{2kT}}
=\frac{\Omega}{\vert J\vert}.
\label{2.02}
\end{eqnarray}
Moreover,
the temperature-driven continuous phase transition
in the square-lattice Ising model
corresponds to the transverse field-driven continuous phase transition
at $T=0$
in the transverse Ising chain.
The critical temperature $T_c$
corresponds
to the critical transverse field $\Omega_c$.
The dependences such as
the Helmholtz free energy,
entropy,
and specific heat
against temperature $T$
correspond to
the dependences such as
the ground-state energy,
transverse magnetization,
and static transverse susceptibility
against transverse field $\Omega$.
Finally,
the correlation length $\xi$
corresponds to the inverse energy gap $\Delta$,
$\xi\sim \Delta^{-1}$.
The behavior of some ground-state quantities
found by P.~Pfeuty\cite{011}
is reported in Table \ref{tab01} and Fig. \ref{fig02}.
\begin{table}
\tbl{Towards the correspondence
between
thermodynamic properties of the square-lattice Ising model
and
the ground-state properties of the transverse Ising chain.}
{\tabcolsep10pt
\begin{tabular}{|c|c|}                                                   \hline
square-lattice Ising model &
transverse Ising chain                                                \\ \hline
$m^z\sim\left(T_c-T\right)^\frac{1}{8}$ &
$\langle s^x\rangle\sim \left(\Omega_c-\Omega\right)^\frac{1}{8}$     \\ 
$c\sim\ln\vert T-T_c\vert$  &
$\chi^z\sim\ln\vert \Omega-\Omega_c\vert$                             \\ 
$\xi\sim\vert T-T_c\vert^{-1}$ &
$\Delta\sim\vert \Omega-\Omega_c\vert$                                \\ \hline
\end{tabular}}
\label{tab01}
\end{table}

Now I wish to explain briefly
how can the statistical mechanics calculations
for the transverse Ising chain
be carried out
without making any approximation.
In terms of the spin raising and lowering operators
$s_n^\pm=s_n^x\pm{\mbox{i}}s_n^y$
the Hamiltonian of the transverse Ising chain becomes
as follows:
\begin{eqnarray}
H=\sum_n\Omega_n\left(s^+_ns^-_n-\frac{1}{2}\right)
\nonumber\\
+\sum_n\frac{I_n}{2}
\left(s^+_ns^+_{n+1}+s^+_ns^-_{n+1}+s^-_ns^+_{n+1}+s^-_ns^-_{n+1}\right).
\label{2.03}
\end{eqnarray}
Bearing in mind a case of regularly alternating chain
we consider a model more general than (\ref{2.01})
assuming that the Hamiltonian parameters
are site-dependent,
$\Omega\to\Omega_n$,
$J\to J_n=2I_n$.
Although (\ref{2.03}) is a bilinear form
in terms of $s^\pm$ operators,
further calculations appear to be complicated
because of the commutation relations
which are of the Fermi type at the same site
\begin{eqnarray}
\left\{s_n^-,s_n^+\right\}=1,
\;\;\;
\left\{s_n^-,s_n^-\right\}
=\left\{s_n^+,s_n^+\right\}=0
\label{2.04}
\end{eqnarray}
and of the Bose type at different sites
\begin{eqnarray}
\left[s_n^-,s_m^+\right]
=\left[s_n^-,s_m^-\right]
=\left[s_n^+,s_m^+\right]=0,
\;\;\;
n\ne m.
\label{2.05}
\end{eqnarray}
We may use
\index{Jordan-Wigner transformation}
the Jordan-Wigner transformation\footnote
{The Jordan-Wigner transformation
of the spin operators to spinless fermions
is also described in Chapter ? (see Eq. (3) of that Chapter).}
to introduce the Fermi operators
according to the following formulas:
\begin{eqnarray}
c_n=\left(-2s_1^z\right)\left(-2s_2^z\right)
\ldots \left(-2s_{n-1}^z\right)s^-_n,
\label{2.06}
\end{eqnarray}
\begin{eqnarray}
c_n^+=\left(-2s_1^z\right)\left(-2s_2^z\right)
\ldots \left(-2s_{n-1}^z\right)s^+_n.
\label{2.07}
\end{eqnarray}
Really,
the introduced operators always satisfy the Fermi commutation relations
\begin{eqnarray}
\left\{c_n,c_m^+\right\}=\delta_{nm},
\;\;\;
\left\{c_n,c_m\right\}
=\left\{c_n^+,c_m^+\right\}=0.
\label{2.08}
\end{eqnarray}
Moreover,
the Hamiltonian (\ref{2.03}) in terms of the Fermi operators
(\ref{2.06}), (\ref{2.07})
remains to be a bilinear form.
Namely,
\begin{eqnarray}
H=\sum_n\Omega_n\left(c^+_nc_n-\frac{1}{2}\right)
\nonumber\\
+\sum_n\frac{I_n}{2}
\left(c^+_nc^+_{n+1}+c^+_nc_{n+1}-c_nc^+_{n+1}-c_nc_{n+1}\right)
\nonumber\\
=
-\frac{1}{2}\sum_n\Omega_n
+\sum_{n,m}\left(c_n^+A_{nm}c_m
+\frac{1}{2}\left(c_n^+B_{nm}c_m^+-c_nB_{nm}c_m\right)\right)
\label{2.09}
\end{eqnarray}
where
\begin{eqnarray}
A_{nm}
=\Omega_n\delta_{nm}
+\frac{1}{2}I_n\delta_{m,n+1}
+\frac{1}{2}I_{n-1}\delta_{m,n-1}=A_{mn},
\label{2.10}
\end{eqnarray}
\begin{eqnarray}
B_{nm}
=\frac{1}{2}I_n\delta_{m,n+1}
-\frac{1}{2}I_{n-1}\delta_{m,n-1}
=-B_{mn}.
\label{2.11}
\end{eqnarray}
For periodic boundary conditions
imposed on (\ref{2.03}),
the transformed Hamiltonian (\ref{2.09})
should also contain
the so-called boundary term
which is omitted
since we send $N$ to infinity.

To diagonalize a bilinear in Fermi operators form (\ref{2.09})
we perform
the linear canonical transformation
with real coefficients
\begin{eqnarray}
\eta_k
=\sum_n\left(g_{kn}c_n+h_{kn}c_n^+\right),
\nonumber\\
\eta_k^+
=\sum_n\left(g_{kn}c_n^++h_{kn}c_n\right).
\label{2.12}
\end{eqnarray}
The resulting Hamiltonian becomes as follows:
\begin{eqnarray}
H
=\sum_{k=1}^N\Lambda_k
\left(\eta_k^+\eta_k-\frac{1}{2}\right),
\nonumber\\
\left\{\eta_{k^{\prime}},\eta^+_{k^{\prime\prime}}\right\}
=\delta_{k^{\prime}k^{\prime\prime}}
\;\;\;
\left\{\eta_{k^{\prime}},\eta_{k^{\prime\prime}}\right\}
=\left\{\eta^+_{k^{\prime}},\eta^+_{k^{\prime\prime}}\right\}=0
\label{2.13}
\end{eqnarray}
if the coefficients $g_{kn}$, $h_{kn}$
or, more precisely, their linear combinations
\begin{eqnarray}
\Phi_{kn}=g_{kn}+h_{kn},
\nonumber\\
\Psi_{kn}=g_{kn}-h_{kn}
\label{2.14}
\end{eqnarray}
satisfy the set of equations
\begin{eqnarray}
\Lambda_k^2\Phi_{kn}
=\sum_{j}\Phi_{kj}
\left(\left({\bf{A}}-{\bf{B}}\right)\left({\bf{A}}+{\bf{B}}\right)\right)_{jn},
\nonumber\\
\Lambda_k^2\Psi_{kn}
=\sum_{j}\Psi_{kj}
\left(\left({\bf{A}}+{\bf{B}}\right)\left({\bf{A}}-{\bf{B}}\right)\right)_{jn}.
\label{2.15}
\end{eqnarray}
Eqs. (\ref{2.15}) explicitly read
\begin{eqnarray}
\Omega_{n-1}I_{n-1}\Phi_{k,n-1}
+\left(I_{n-1}^2+\Omega_n^2-\Lambda^2_k\right)\Phi_{kn}
+\Omega_nI_n\Phi_{k,n+1}=0,
\nonumber\\
\Omega_{n}I_{n-1}\Psi_{k,n-1}
+\left(I_{n}^2+\Omega_n^2-\Lambda^2_k\right)\Psi_{kn}
+\Omega_{n+1}I_n\Psi_{k,n+1}=0,
\label{2.16}
\end{eqnarray}
with periodic or open boundary conditions implied.
Eqs. (\ref{2.16}), (\ref{2.14})
determine
the coefficients $g_{kn}$, $h_{kn}$  in (\ref{2.12})
and
the elementary excitation energies $\Lambda_k$ in (\ref{2.13}).

For the uniform transverse Ising chain,
$\Omega_n=\Omega$, $I_n=I$,
the bilinear in Fermi operators form (\ref{2.09})
becomes diagonal after performing:
\index{Fourier transformation}
i) the Fourier transformation
\begin{eqnarray}
c_j=\frac{1}{\sqrt{N}}
\sum_\kappa\exp\left(-{\mbox{i}}\kappa j\right)c_\kappa,
\nonumber\\
c_j^+=\frac{1}{\sqrt{N}}
\sum_\kappa\exp\left({\mbox{i}}\kappa j\right)c_\kappa^+
\label{2.17}
\end{eqnarray}
with
$\kappa=\frac{2\pi}{N}n$
and
$n=-\frac{N}{2},-\frac{N}{2}+1,\ldots,\frac{N}{2}-1$
if $N$ is even
or
$n=-\frac{N-1}{2},-\frac{N-1}{2}+1,\ldots,\frac{N-1}{2}$
if $N$ is odd
and
\index{Bogolyubov transformation}
ii) the Bogolyubov transformation
\begin{eqnarray}
\eta_\kappa=x_\kappa c_\kappa +y_\kappa c_{-\kappa}^+,
\nonumber\\
\eta_{-\kappa}^+=y_{-\kappa}^* c_\kappa +x_{-\kappa}^* c_{-\kappa}^+
\label{2.18}
\end{eqnarray}
with
\begin{eqnarray}
x_\kappa
=\frac{{\mbox{i}}I\sin\kappa}
{\sqrt{2\Lambda_\kappa\left(\Lambda_\kappa-\epsilon_\kappa\right)}},
\nonumber\\
y_\kappa
=\sqrt{\frac{\Lambda_\kappa-\epsilon_\kappa}{2\Lambda_\kappa}}
\label{2.19}
\end{eqnarray}
and
\begin{eqnarray}
\Lambda_\kappa
=\sqrt{\epsilon^2_\kappa+I^2\sin^2\kappa},
\nonumber\\
\epsilon_\kappa=\Omega+I\cos\kappa.
\label{2.20}
\end{eqnarray}

Knowing the energies of noninteracting fermions
$\Lambda_k$
in (\ref{2.13})
we immediately obtain
the Helmholtz free energy per site
(and hence all thermodynamic quantities),
\begin{eqnarray}
f=-\frac{1}{N\beta}\ln{\mbox{Tr}}\exp\left(-\beta H\right)
\nonumber\\
=-\frac{1}{N\beta}\ln\prod_{k=1}^N
\left(\exp\left(\frac{\beta\Lambda_k}{2}\right)
+\exp\left(-\frac{\beta\Lambda_k}{2}\right)\right)
\nonumber\\
=-\frac{1}{N\beta}\sum_{k=1}^N\ln\left(2\cosh\frac{\beta\Lambda_k}{2}\right)
\nonumber\\
=-\frac{1}{2\pi\beta}\int_{-\pi}^\pi
{\mbox{d}}\kappa\ln\left(2\cosh\frac{\beta\Lambda_\kappa}{2}\right)
\label{2.21}
\end{eqnarray}
(the last line in (\ref{2.21}) refers to the uniform case (\ref{2.20})).
Note,
that we can obtain all thermodynamic quantities
knowing the distributions
of the energies of elementary excitations,
\begin{eqnarray}
\rho(E)=\frac{1}{N}\sum_{k=1}^N\delta\left(E-\Lambda_k\right),
\nonumber\\
\int_{-\infty}^\infty{\mbox{d}}E\rho(E)=1,
\label{2.22}
\end{eqnarray}
or the distribution
of the squared energies of elementary excitations
\begin{eqnarray}
R(E^2)=\frac{1}{N}\sum_{k=1}^N\delta\left(E^2-\Lambda^2_k\right),
\nonumber\\
\int_{0}^\infty{\mbox{d}}{E^2}R(E^2)=1.
\label{2.23}
\end{eqnarray}
Really,
the density of states $\rho(E)$ (\ref{2.22})
or
the density of states $R(E^2)$ (\ref{2.23})
immediately yields
\begin{eqnarray}
f=-\frac{1}{\beta}\int_{-\infty}^\infty{\mbox{d}}E
\rho(E)\ln \left(2\cosh\frac{\beta E}{2}\right)
\nonumber\\
=-\frac{2}{\beta}\int_{0}^\infty{\mbox{d}}E E
R(E^2)\ln \left(2\cosh\frac{\beta E}{2}\right).
\label{2.24}
\end{eqnarray}

The calculation of the spin correlation functions
in the fermionic picture
looks as follows.
First we write down
the relations
between the spin and Fermi operators.
We have
\begin{eqnarray}
s_n^z
=c_n^+c_n-\frac{1}{2}
=-\frac{1}{2}\left(c_n^++c_n\right)\left(c^+_n-c_n\right)
=-\frac{1}{2}\varphi^+_n\varphi_n^-
\label{2.25}
\end{eqnarray}
where we have introduced the operators
$\varphi_n^\pm=c_n^+\pm c_n$\footnote
{The operators $\varphi^\pm$
are related to the Clifford operators $\Gamma^1$, $\Gamma^2$
(Chapter ?, Eq. (5))
as follows:
$\varphi_n^+=\Gamma_n^1$,
$\varphi_n^-=-{\mbox{i}}\Gamma_n^2$.}.
Further,
\begin{eqnarray}
s_n^x
=\frac{1}{2}\varphi^+_1\varphi^-_1\ldots\varphi^+_{n-1}\varphi^-_{n-1}
\varphi_n^+,
\label{2.26}
\end{eqnarray}
\begin{eqnarray}
s_n^y
=\frac{1}{2{\mbox{i}}}\varphi^+_1\varphi^-_1\ldots\varphi^+_{n-1}\varphi^-_{n-1}
\varphi_n^-.
\label{2.27}
\end{eqnarray}
Note that the relations between the operators $s_n^x$, $s_n^y$
and the Fermi operators are nonlocal
since the r.h.s. in Eqs. (\ref{2.26}), (\ref{2.27})
involve the Fermi operators attached to all previous sites.
That is in contrast to the similar relation (\ref{2.25})
for the operator $s_n^z$,
the r.h.s. of which contains only the Fermi operators
attached to the same site $n$.

Now we can rewrite the spin correlation functions in fermionic language.
For example,
\begin{eqnarray}
\langle s_n^z s_{n+m}^z \rangle
=\frac{1}{4}\langle\varphi_n^+\varphi_n^-\varphi_{n+m}^+\varphi_{n+m}^-\rangle,
\label{2.28}
\end{eqnarray}
\begin{eqnarray}
\langle s_n^x s_{n+m}^x \rangle
=\frac{1}{4}\langle\varphi_n^-\varphi_{n+1}^+\varphi_{n+1}^-\ldots
\varphi_{n+m-1}^+\varphi_{n+m-1}^-\varphi_{n+m}^+\rangle.
\label{2.29}
\end{eqnarray}
To get the latter result we use the relations
\begin{eqnarray}
\left\{\varphi_n^+,\varphi_m^+\right\}
=-\left\{\varphi_n^-,\varphi_m^-\right\}
=2\delta_{nm},
\;\;\;
\left\{\varphi_n^+,\varphi_m^-\right\}=0.
\label{2.30}
\end{eqnarray}

The operators $\varphi_n^\pm$ are linear combinations
of operators $\eta_k$, $\eta^+_k$
involved into (\ref{2.13}),
\begin{eqnarray}
\varphi_n^+
=\sum_{p=1}^N\Phi_{pn}
\left(\eta_p^++\eta_p\right),
\nonumber\\
\varphi_n^-
=\sum_{p=1}^N\Psi_{pn}
\left(\eta_p^+-\eta_p\right),
\label{2.31}
\end{eqnarray}
or for the uniform chain
\begin{eqnarray}
\varphi_n^+
=\frac{1}{\sqrt{N}}
\sum_{\kappa}\exp\left({\mbox{i}}\kappa n\right)\left(x_\kappa+y_\kappa\right)
\left(\eta_\kappa^++\eta_{-\kappa}\right),
\nonumber\\
\varphi_n^+
=\frac{1}{\sqrt{N}}
\sum_{\kappa}\exp\left({\mbox{i}}\kappa n\right)\left(x_\kappa-y_\kappa\right)
\left(\eta_\kappa^+-\eta_{-\kappa}\right).
\label{2.32}
\end{eqnarray}
Therefore,
we calculate (\ref{2.28}), (\ref{2.29})
\index{Wick-Bloch-de Dominicis theorem}
using the Wick-Bloch-de Dominicis theorem\footnote
{See also Section 3.1.2 in Chapter ?.}.
Namely,
\begin{eqnarray}
4\langle s_n^z s_{n+m}^z \rangle
=\langle\varphi_n^+\varphi_n^-\varphi_{n+m}^+\varphi_{n+m}^-\rangle
\nonumber\\
=\langle\varphi_n^+\varphi_n^-\rangle\langle\varphi_{n+m}^+\varphi_{n+m}^-\rangle
\nonumber\\
-\langle\varphi_n^+\varphi_{n+m}^+\rangle\langle\varphi_{n}^-\varphi_{n+m}^-\rangle
+\langle\varphi_n^+\varphi_{n+m}^-\rangle\langle\varphi_{n}^-\varphi_{n+m}^+\rangle.
\label{2.33}
\end{eqnarray}
The r.h.s. of Eq. (\ref{2.33})
may be compactly written
\index{Pfaffian}
as the Pfaffian
of the $4\times 4$ antisymmetric matrix
\begin{eqnarray}
4\langle s_n^z s_{n+m}^z \rangle
\nonumber\\
={\mbox{Pf}}
\left(
\begin{array}{cccc}
0 &
\langle \varphi^+_n\varphi_n^- \rangle &
\langle \varphi^+_n\varphi_{n+m}^+ \rangle &
\langle \varphi^+_n\varphi_{n+m}^- \rangle \\
-\langle \varphi^+_n\varphi_n^- \rangle &
0 &
\langle \varphi^-_n\varphi_{n+m}^+ \rangle &
\langle \varphi^-_n\varphi_{n+m}^- \rangle \\
-\langle \varphi^+_n\varphi_{n+m}^+ \rangle &
-\langle \varphi^-_n\varphi_{n+m}^+ \rangle &
0 &
\langle \varphi^+_{n+m}\varphi_{n+m}^- \rangle \\
-\langle \varphi^+_n\varphi_{n+m}^- \rangle &
-\langle \varphi^-_n\varphi_{n+m}^- \rangle &
-\langle \varphi^+_{n+m}\varphi_{n+m}^- \rangle &
0
\end{array}
\right).
\label{2.34}
\end{eqnarray}
Similarly,
\begin{eqnarray}
4\langle s_n^x s_{n+m}^x \rangle
\nonumber\\
={\mbox{Pf}}
\left(
\begin{array}{ccccc}
0                                          &
\langle \varphi^-_n\varphi_{n+1}^+ \rangle &
\langle \varphi^-_n\varphi_{n+1}^- \rangle &
\ldots                                     &
\langle \varphi^-_n\varphi_{n+m}^+ \rangle \\
-\langle \varphi^-_n\varphi_{n+1}^+ \rangle    &
0                                              &
\langle \varphi^+_{n+1}\varphi_{n+1}^- \rangle &
\ldots                                         &
\langle \varphi^+_{n+1}\varphi_{n+m}^+ \rangle \\
-\langle \varphi^-_n\varphi_{n+1}^- \rangle     &
-\langle \varphi^+_{n+1}\varphi_{n+1}^- \rangle &
0                                               &
\ldots                                          &
\langle \varphi^-_{n+1}\varphi_{n+m}^+ \rangle  \\
\vdots &
\vdots &
\vdots &
\cdots &
\vdots \\
-\langle \varphi^-_n\varphi_{n+m}^+ \rangle     &
-\langle \varphi^+_{n+1}\varphi_{n+m}^+ \rangle &
-\langle \varphi^-_{n+1}\varphi_{n+m}^+ \rangle &
\ldots                                          &
0
\end{array}
\right).
\nonumber\\
\label{2.35}
\end{eqnarray}
The elementary contractions
involved into (\ref{2.34}), (\ref{2.35})
read
\begin{eqnarray}
\langle\varphi_n^+\varphi_{m}^+\rangle
=\sum_{k=1}^N\Phi_{kn}\Phi_{km}=\delta_{nm},
\nonumber\\
\langle\varphi_n^+\varphi_{m}^-\rangle
=\sum_{k=1}^N\Phi_{kn}\Psi_{km}\tanh\frac{\beta\Lambda_p}{2},
\nonumber\\
\langle\varphi_n^-\varphi_{m}^+\rangle
=-\sum_{k=1}^N\Psi_{kn}\Phi_{km}\tanh\frac{\beta\Lambda_p}{2},
\nonumber\\
\langle\varphi_n^-\varphi_{m}^-\rangle
=-\sum_{k=1}^N\Psi_{kn}\Psi_{km}=-\delta_{nm}.
\label{2.36}
\end{eqnarray}
In the uniform case,
instead of (\ref{2.36}) we have
\begin{eqnarray}
\langle\varphi_n^+\varphi_{m}^+\rangle
=-\langle\varphi_n^-\varphi_{m}^-\rangle=\delta_{nm},
\nonumber\\
\langle\varphi_n^+\varphi_{m}^-\rangle
=\frac{1}{N}\sum_\kappa
\exp\left({\mbox{i}}\kappa\left(n-m\right)\right)
\frac{\Omega+I\exp\left(-{\mbox{i}}\kappa\right)}{\Lambda_\kappa}
\tanh\frac{\beta\Lambda_\kappa}{2},
\nonumber\\
\langle\varphi_n^-\varphi_{m}^+\rangle
=-\frac{1}{N}\sum_\kappa
\exp\left(-{\mbox{i}}\kappa\left(n-m\right)\right)
\frac{\Omega+I\exp\left(-{\mbox{i}}\kappa\right)}{\Lambda_\kappa}
\tanh\frac{\beta\Lambda_\kappa}{2}.
\label{2.37}
\end{eqnarray}

It is worthy to recall some properties of the Pfaffians
which are used in calculating them.
In the first numerical studies
the authors used the relation
\begin{eqnarray}
\left({\mbox{Pf}}{\bf{A}}\right)^2
={\mbox{det}}{\bf{A}}
\label{2.38}
\end{eqnarray}
and computed numerically the determinants
which gave the values of Pfaffians.
On the other hand,
the Pfaffian may be computed directly,\cite{016,017}
noting that
\begin{eqnarray}
{\mbox{Pf}}\left({\bf{U}}^T{\bf{A}}{\bf{U}}\right)
={\mbox{det}}{\bf{U}}\;{\mbox{Pf}}{\bf{A}}
\label{2.39}
\end{eqnarray}
and that
\begin{eqnarray}
{\mbox{Pf}}
\left(
\begin{array}{cccccc}
0      &
R_{12} &
0      &
0      &
\ldots &
0      \\
-R_{12} &
0       &
0       &
0       &
\ldots  &
0       \\
0         &
0         &
0         &
R_{34}    &
\ldots    &
0         \\
0       &
0       &
-R_{34} &
0       &
\ldots  &
0       \\
\vdots &
\vdots &
\vdots &
\vdots &
\ldots &
\vdots \\
0 &
0 &
0 &
0 &
\ldots &
0
\end{array}
\right)
=R_{12}R_{34}\ldots .
\label{2.40}
\end{eqnarray}

Let me finally show how the results
for the transverse Ising chain obtained by P.~Pfeuty\cite{011}
arise in the described approach.
Recalling Eq. (\ref{2.13})
(or Eq. (\ref{2.21}))
it is not hard to see
that the ground-state energy per site can be written as
\begin{eqnarray}
e_0
=\frac{1}{2\pi}
\int_{-\pi}^{\pi}{\mbox{d}}\kappa\left(-\frac{\Lambda_\kappa}{2}\right)
\nonumber\\
=-\frac{1}{4\pi}
\int_{-\pi}^{\pi}{\mbox{d}}\kappa\sqrt{\Omega^2+2\Omega I\cos\kappa +I^2}
\nonumber\\
=-\frac{I}{4\pi}
\int_{-\pi}^{\pi}{\mbox{d}}\kappa\sqrt{\lambda^2+2\lambda\cos\kappa+1}
\nonumber\\
=-\frac{I\left(1+\lambda\right)}{\pi}
\int_{0}^{\frac{\pi}{2}}{\mbox{d}}\phi
\sqrt{1-\left(\frac{2\sqrt{\lambda}}{1+\lambda}\right)^2 \sin^2\phi}
\nonumber\\
=-\frac{I\left(1+\lambda\right)}{\pi}
E\left(\frac{\pi}{2},\frac{2\sqrt{\lambda}}{1+\lambda}\right)
\label{2.41}
\end{eqnarray}
where
$\lambda=\frac{\Omega}{I}$
and
$$E\left(\frac{\pi}{2},a\right)
=\int_0^\frac{\pi}{2}
{\mbox{d}}\phi\sqrt{1-a^2\sin^2\phi}
$$
is the complete elliptic integral of the second kind.\cite{018}
In the vicinity of $a=1$
\begin{eqnarray}
E\left(\frac{\pi}{2}, a\right)
\sim
1+\frac{1-a^2}{4}\ln\frac{16}{1-a^2}
\label{2.41_2}
\end{eqnarray}
(see Ref. \refcite{018_19,003_4})
and therefore
in the vicinity of $\lambda_c=1$
the ground-state energy (\ref{2.41})
contains the nonanalytic contribution
\begin{eqnarray}
e_0
\sim\left(\lambda-\lambda_c\right)^2
\ln\left\vert\lambda-\lambda_c\right\vert.
\label{2.42}
\end{eqnarray}
As a result
the ground-state transverse magnetization contains
the nonanalytic contribution
\begin{eqnarray}
m^z
\sim\left(\lambda-\lambda_c\right)
\ln\left\vert\lambda-\lambda_c\right\vert
\label{2.43}
\end{eqnarray}
and
the ground-state static transverse susceptibility
exhibits a logarithmic singularity
\begin{eqnarray}
\chi^z
\sim\ln\left\vert\lambda-\lambda_c\right\vert.
\label{2.44}
\end{eqnarray}
At $\lambda_c=1$
the energy spectrum is gapless,
i.e. there is a zero-energy elementary excitation.
In the vicinity of $\lambda_c=1$
the energy gap is given by the smallest value of
$\Lambda_\kappa=\sqrt{\Omega^2+2\Omega I\cos\kappa+I^2}$
and hence
\begin{eqnarray}
\Delta
\sim\left\vert\lambda-\lambda_c\right\vert.
\label{2.45}
\end{eqnarray}

\begin{figure}[th]
\centerline{\psfig{file=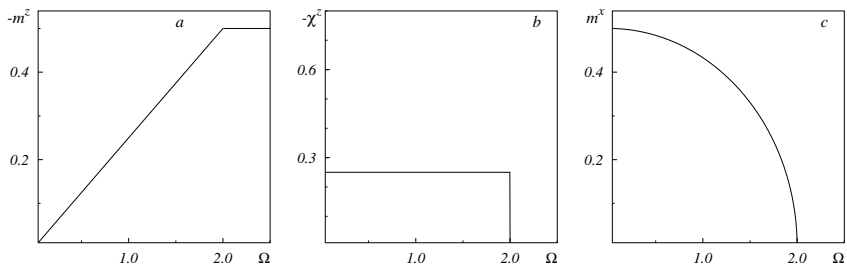,width=4.6in,angle=0}}
\vspace*{8pt}
\caption{The same as in Fig.~\ref{fig02}
for the classical transverse Ising chain (\ref{2.46})
($s=\frac{1}{2}$, $I=-1$).}
\label{fig03}
\end{figure}

To end up
the discussion of the ground-state properties of the transverse Ising chain,
let me emphasize that the quantum phase transition
appears since the spin variables are the q-numbers
rather than the c-numbers.
To demonstrate this explicitly
we may consider spin-vector (instead of spin-matrix) transverse Ising chain.
The model consists of $N\to\infty$ 3-componet vectors
$${\bf{s}}
=(s^x,s^y,s^z)
=(s\sin\theta\cos\phi,s\sin\theta\sin\phi,s\cos\theta)$$
which are governed by the Hamiltonian
\begin{eqnarray}
H
=\sum_n\Omega s\cos\theta_n
+\sum_n 2Is^2\sin\theta_n\sin\theta_{n+1}\cos\phi_n\cos\phi_{n+1}.
\label{2.46}
\end{eqnarray}
Here $s$ is the value of the spin
which plays only a quantitative role
and further is put $s=\frac{1}{2}$.
To obtain the ground-state energy
we should place all spins in $xz$ plane
putting $\phi_n=0$ ($I<0$).
Moreover,
the angles $\theta_n$
must minimize the sum of contributions
due to the interaction with the field
and
due to the intersite interaction.
Thus, the ground-state energy ansatz reads
\begin{eqnarray}
E_0(\theta)
=\frac{1}{2}N\Omega \cos\theta
-\frac{1}{2}N\vert I\vert \sin^2\theta
\label{2.47}
\end{eqnarray}
where $\theta$ is determined to minimize
$E_0(\theta)$ (\ref{2.47}).
Obviously
\begin{eqnarray}
\cos\theta
=\left\{
\begin{array}{rl}
1, &
{\mbox{if}}\;\;\;
\Omega<-2\vert I\vert,\\
-\frac{\Omega}{2\vert I\vert}, &
{\mbox{if}}\;\;\;
-2\vert I\vert\le \Omega<2\vert I\vert,\\
-1, &
{\mbox{if}}\;\;\;
2\vert I\vert\le \Omega.
\end{array}
\right.
\label{2.48}
\end{eqnarray}
As a result,
we find that the ground-state energy per site
is given by
\begin{eqnarray}
e_0
=\left\{
\begin{array}{rl}
\frac{\Omega}{2}, &
{\mbox{if}}\;\;\;
\Omega<-2\vert I\vert,\\
-\frac{\vert I\vert}{2}
\left(
1+\frac{\Omega^2}{4 I^2}\right), &
{\mbox{if}}\;\;\;
-2\vert I\vert\le \Omega<2\vert I\vert,\\
-\frac{\Omega}{2}, &
{\mbox{if}}\;\;\;
2\vert I\vert\le \Omega.
\end{array}
\right.
\label{2.49}
\end{eqnarray}
Moreover, the $x$- and $z$-magnetizations
which can be obtained after inserting (\ref{2.48}) into
the formulas
$m^x=s\sin\theta$
and
$m^z=s\cos\theta$,
respectively,
behave as
\begin{eqnarray}
m^x
=\left\{
\begin{array}{rl}
0, &
{\mbox{if}}\;\;\;
\Omega<-2\vert I\vert,\\
\frac{1}{2}\sqrt{1-\frac{\Omega^2}{4I^2}}, &
{\mbox{if}}\;\;\;
-2\vert I\vert\le \Omega<2\vert I\vert,\\
0, &
{\mbox{if}}\;\;\;
2\vert I\vert\le\Omega
\end{array}
\right.
\label{2.50}
\end{eqnarray}
and
\begin{eqnarray}
m^z
=\left\{
\begin{array}{rl}
\frac{1}{2}, &
{\mbox{if}}\;\;\;
\Omega<-2\vert I\vert,\\
-\frac{\Omega}{4\vert I\vert}, &
{\mbox{if}}\;\;\;
-2\vert I\vert\le \Omega<2\vert I\vert,\\
-\frac{1}{2}, &
{\mbox{if}}\;\;\;
2\vert I\vert\le\Omega,
\end{array}
\right.
\label{2.51}
\end{eqnarray}
respectively.
Evidently,
there is no singularity in the ground-state dependence
of the static transverse susceptibility
$\chi^z=\frac{\partial m^z}{\partial\Omega}$
on the transverse field $\Omega$.
As the transverse field increases from zero to $2\vert I\vert$,
the on-site magnetic moments
smoothly change their direction from
$x$- to $z$-axis.
The quantum phase transition appears only due to the quantum fluctuations
which
arise when the spin components do not commute
and
no phase transition occurs
when the spin components are classical variables
(compare Fig.~\ref{fig02} and Fig.~\ref{fig03}).

\section{Spin-$\frac{1}{2}$ Ising Chain in a Transverse Field
         with Regularly Alternating Hamiltonian Parameters:
         Continued Fraction Approach
         \label{secd3}}

We are turning to a discussion
of the effects
of regularly alternating exchange interactions and fields
on the quantum phase transition
inherent in the transverse Ising chain.
Note,
that there was a great deal of work done
examining the effects of different modifications of the skeleton model
on the quantum phase transition.
Thus, J.~M.~Luck\cite{019}
analyzed the critical behavior
of the chain with an aperiodic sequence of interactions,
D.~S.~Fisher\cite{020}
performed an extensive real-space renormalization group study
of random chains,
F.~Igl\'{o}i et al\cite{021} reported
a renormalization-group study of aperiodic chains.
However, a simpler case of regularly alternating transverse Ising chain
still contains a good deal of unexplored physics.
It appears possible
to obtain analytically
exact results for thermodynamic quantities of such  models
using the Jordan-Wigner fermionization
and the continued fraction approach.\cite{022,023}
Moreover,
for certain sets of the Hamiltonian parameters
the exact results may be obtained
for the spin correlation functions.\cite{024}
In all other cases,
spin correlations may be explored numerically
at a high precision level
considering sufficiently long chains
up to a few thousand sites.\cite{024,025}
It is appropriate
to mention here
the old studies\cite{026,027}
referring to the chains of period 2
which were extended further
to the one-dimensional anisotropic $XY$ model
on superlattices.\cite{028,029}
The quantum critical points
in the anisotropic $XY$ chains in a transverse field
with periodically varying intersite interactions
(having periods 2 and 3)
have been determined recently\cite{030}
using the transfer matrix method.

I begin with explaining
how the thermodynamic quantities
can be derived using continued fractions.
For this purpose let me recall
that the energies of the Jordan-Wigner fermions
which determine the thermodynamic properties of a spin chain
are involved into Eqs. (\ref{2.16}).
Consider, for example, the first set of equations in (\ref{2.16})
which can be rewritten in the matrix form as follows:
\begin{eqnarray}
\label{3.01}
\left({\bf{H}}-\Lambda_k^2{\bf{1}}\right)
\left(
\begin{array}{c}
\Phi_{k1}\\
\Phi_{k2}\\
\Phi_{k3}\\
\vdots\\
\Phi_{kN}
\end{array}
\right)
=0
\end{eqnarray}
where
\begin{eqnarray}
{\bf{H}}=
\left(
\begin{array}{ccccccc}
\vdots &
\vdots &
\vdots &
\vdots &
\vdots &
\vdots &
\vdots \\
\ldots &
\Omega_{n-2}I_{n-2} &
I_{n-2}^2+\Omega_{n-1}^2 &
\Omega_{n-1}I_{n-1} &
0 &
0 &
\ldots \\
\ldots &
0 &
\Omega_{n-1} I_{n-1} &
I_{n-1}^2+\Omega_n^2 &
\Omega_n I_n &
0 &
\ldots \\
\ldots &
0 &
0 &
\Omega_nI_n &
I_n^2 +\Omega_{n+1}^2 &
\Omega_{n+1}I_{n+1} &
\ldots \\
\vdots &
\vdots &
\vdots &
\vdots &
\vdots &
\vdots &
\vdots
\end{array}
\right)
\nonumber\\
\label{3.02}
\end{eqnarray}
and ${\bf{1}}$ denotes the unit matrix.
Thus, all $\Lambda_k^2$ are the eigenvalues of the $N\times N$ matrix
(\ref{3.02})
entering (\ref{3.01}).
To find  their distribution $R(E^2)$ (\ref{2.23})
we may use
\index{Green function approach}
the Green function approach.
Consider a matrix ${\bf{G}}$
composed of the elements
$G_{nm}=G_{nm}(E^2)$
which is introduced by the equation
\begin{eqnarray}
\left(E^2{\bf{1}}-{\bf{H}}\right){\bf{G}}={\bf{1}}.
\label{3.03}
\end{eqnarray}
Having composed a matrix ${\bf{U}}^+$
of the eigenvectors of the matrix ${\bf{H}}$ (\ref{3.02}),
i.e.
\begin{eqnarray}
{\bf{U}}{\bf{H}}{\bf{U}}^+
=
\left(
\begin{array}{cccc}
\Lambda_1^2 &
0 &
\ldots &
0 \\
0 &
\Lambda_2^2 &
\ldots &
0\\
\vdots &
\vdots &
\vdots &
\vdots \\
0 &
0 &
\ldots &
\Lambda_N^2
\end{array}
\right),
\label{3.04}
\end{eqnarray}
one finds that
\begin{eqnarray}
\left(
\begin{array}{cccc}
E^2-\Lambda_1^2 &
0 &
\ldots &
0 \\
0 &
E^2-\Lambda_2^2 &
\ldots &
0 \\
\vdots &
\vdots &
\vdots &
\vdots \\
0 &
0 &
\ldots &
E^2-\Lambda_N^2
\end{array}
\right)
{\bf{U}}
{\bf{G}}
{\bf{U}}^+
={\bf{1}}.
\label{3.05}
\end{eqnarray}
As a result
\begin{eqnarray}
{\mbox{Tr}}{\bf{G}}
=\sum_{k=1}^N\left({\bf{U}}{\bf{G}}{\bf{U}}^+\right)_{kk}
=\sum_{k=1}^N\frac{1}{E^2-\Lambda_k^2}.
\label{3.06}
\end{eqnarray}
Making the substitution
\begin{eqnarray}
E^2\to E^2\pm {\mbox{i}}\epsilon,
\;\;\;
\epsilon\to +0
\label{3.07}
\end{eqnarray}
and using the symbolic identity
\begin{eqnarray}
\frac{1}{E^2-\Lambda_k^2\pm {\mbox{i}}\epsilon}
={\cal{P}}\frac{1}{E^2-\Lambda_k^2}
\mp{\mbox{i}}\pi\delta\left(E^2-\Lambda_k^2\right),
\;\;\;
\epsilon\to +0
\label{3.08}
\end{eqnarray}
one arrives at the relation
\begin{eqnarray}
\mp\frac{1}{N\pi}\sum_{n=1}^N\Im G_{nn}(E^2\pm{\mbox{i}}\epsilon)
=\frac{1}{N}\sum_{n=1}^N\delta(E^2-\Lambda_n^2)=R(E^2).
\label{3.09}
\end{eqnarray}
Thus,
our task is to find the diagonal Green functions $G_{nn}$
defined by Eqs. (\ref{3.03}), (\ref{3.02}).

Considering the equation for $G_{nn}$,
\begin{eqnarray}
\left(
E^2-\Omega_n^2-I_{n-1}^2
\right)
G_{nn}
-\Omega_{n-1}I_{n-1}G_{n-1,n}
-\Omega_nI_nG_{n+1,n}
=1,
\label{3.10}
\end{eqnarray}
and then
the equations
\begin{eqnarray}
\left(
E^2-\Omega_{n-1}^2-I_{n-2}^2
\right)
G_{n-1,n}
-\Omega_{n-2}I_{n-2}G_{n-2,n}
-\Omega_{n-1}I_{n-1}G_{nn}
=0,
\nonumber\\
\left(
E^2-\Omega_{n-2}^2-I_{n-3}^2
\right)
G_{n-2,n}
-\Omega_{n-3}I_{n-3}G_{n-3,n}
-\Omega_{n-2}I_{n-2}G_{n-1,n}
=0
\nonumber\\
\label{3.11}
\end{eqnarray}
etc.
to determine
$\frac{G_{n-1,n}}{G_{nn}}$,
$\frac{G_{n-2,n}}{G_{n-1,n}}$
etc.,
and
the equations
\begin{eqnarray}
\left(
E^2-\Omega_{n+1}^2-I_{n}^2
\right)
G_{n+1,n}
-\Omega_{n}I_{n}G_{nn}
-\Omega_{n+1}I_{n+1}G_{n+2,n}
=0,
\nonumber\\
\left(
E^2-\Omega_{n+2}^2-I_{n+1}^2
\right)
G_{n+2,n}
-\Omega_{n+1}I_{n+1}G_{n+1,n}
-\Omega_{n+2}I_{n+2}G_{n+3,n}
=0
\nonumber\\
\label{3.12}
\end{eqnarray}
etc.
to determine
$\frac{G_{n+1,n}}{G_{nn}}$,
$\frac{G_{n+2,n}}{G_{n+1,n}}$
etc.,
we easily find
\index{continued fractions}
the following continued fraction representation
for the diagonal Green functions
\begin{eqnarray}
G_{nn}=\frac{1}{E^2-\Omega_n^2-I_{n-1}^2-\Delta_n^--\Delta_n^+},
\nonumber\\
\Delta_n^-
=\frac{\Omega_{n-1}^2I_{n-1}^2}{E^2-\Omega_{n-1}^2-I_{n-2}^2
-\frac{\Omega_{n-2}^2I_{n-2}^2}{E^2-\Omega_{n-2}^2-I_{n-3}^2-_{\ddots}}},
\nonumber\\
\Delta_n^+
=\frac{\Omega_{n}^2I_{n}^2}{E^2-\Omega_{n+1}^2-I_{n}^2
-\frac{\Omega_{n+1}^2I_{n+1}^2}{E^2-\Omega_{n+2}^2-I_{n+1}^2-_{\ddots}}}.
\label{3.13}
\end{eqnarray}

A crucial simplification occurs
if a sequence of the Hamiltonian parameters is periodic,
\begin{eqnarray}
\Omega_1I_1\Omega_2I_2\ldots\Omega_pI_p
\Omega_1I_1\Omega_2I_2\ldots\Omega_pI_p\ldots .
\nonumber
\end{eqnarray}
Then
the continued fractions in (\ref{3.13}) become periodic
and can be evaluated exactly
by solving a square equation.
Consider,
for example,
the case $p=1$,
i.e. the uniform chain.
Then $\Delta_n^-=\Delta_n^-=\Delta$
(do not confuse with the energy gap $\Delta$)
and $\Delta$
satisfies the following equation
\begin{eqnarray}
\Delta
=\frac{\Omega^2I^2}{E^2-\Omega^2-I^2-\Delta}.
\label{3.14}
\end{eqnarray}
The solution of Eq. (\ref{3.14}),
\begin{eqnarray}
\Delta
=\frac{1}{2}
\left(
E^2-\Omega^2-I^2
\pm\sqrt{(E^2-\Omega^2-I^2)^2-4\Omega^2I^2}
\right),
\label{3.15}
\end{eqnarray}
yields
\begin{eqnarray}
G_{nn}
=\mp\frac{1}{\sqrt{\left(E^2-\Omega^2-I^2\right)^2-4\Omega^2I^2}}
\label{3.16}
\end{eqnarray}
and hence in accordance with (\ref{3.09})
\begin{eqnarray}
R(E^2)
=
\left\{
\begin{array}{ll}
\frac{1}{\pi\sqrt{4\Omega^2I^2-\left(E^2-\Omega^2-I^2\right)^2}}, &
{\mbox{if}}\;\;\;
\left(\Omega-I\right)^2<E^2<\left(\Omega+I\right)^2, \\
0, &
{\mbox{otherwise}}.
\end{array}
\right.
\label{3.17}
\end{eqnarray}

We may easily repeat the calculations for $p=2$ or $p=3$.
The desired density of states in these cases
is given by
\begin{eqnarray}
R(E^2)
=
\left\{
\begin{array}{ll}
\frac{\vert{\cal{Z}}_{p-1}(E^2)\vert}{p\pi\sqrt{{\cal{A}}_{2p}(E^2)}}, &
{\mbox{if}}\;\;\;
{\cal{A}}_{2p}(E^2)>0, \\
0, &
{\mbox{otherwise}},
\end{array}
\right.
\label{3.18}
\end{eqnarray}
where
${\cal{Z}}_{p-1}(E^2)$
and
${\cal{A}}_{2p}(E^2)
=-\prod_{j=1}^{2p}\left(E^2-a_j\right)$
are polynomials of the order $p-1$  and $2p$,
respectively,
and $a_j\ge 0$ are the roots of ${\cal{A}}_{2p}(E^2)$.
Explicitly,
\begin{eqnarray}
{\cal{Z}}_1(E^2)
=2E^2-\Omega_1^2-\Omega_2^2-I_1^2-I_2^2,
\nonumber\\
{\cal{A}}_4(E^2)
=
4\Omega_1^2\Omega_2^2I_1^2I_2^2
\nonumber\\
-\left(
E^4
-\left(\Omega_1^2+\Omega_2^2+I_1^2+I_2^2\right)E^2
+\Omega_1^2\Omega_2^2+I_1^2I_2^2
\right)^2
\nonumber\\
=-(E^2-a_1)(E^2-a_2)(E^2-a_3)(E^2-a_4),
\nonumber\\
\left\{a_j\right\}
=\left\{
\frac{1}{2}
\left(
\Omega_1^2+\Omega_2^2+I_1^2+I_2^2
\right.
\right.
\nonumber\\
\left.
\left.
\pm\sqrt{(\Omega_1^2+\Omega_2^2+I_1^2+I_2^2)^2
-4(\Omega_1\Omega_2\pm I_1I_2)^2}
\right)
\right\};
\label{3.19}
\end{eqnarray}
\begin{eqnarray}
{\cal{Z}}_{2}(E^2)
=3E^4
-2\left(
\Omega_1^2+\Omega_2^2+\Omega_3^2+I_1^2+I_2^2+I_3^2
\right)E^2
\nonumber\\
+\Omega_1^2\Omega_2^2+\Omega_1^2I_2^2+I_1^2I_2^2
+\Omega_2^2\Omega_3^2+\Omega_2^2I_3^2+I_2^2I_3^2
\nonumber\\
+\Omega_3^2\Omega_1^2+\Omega_3^2I_1^2+I_3^2I_1^2,
\nonumber\\
{\cal{A}}_6(E^2)
=4\Omega_1^2\Omega_2^2\Omega_3^2I_1^2I_2^2I_3^2
\nonumber\\
-\left(
E^6
-\left(\Omega_1^2+\Omega_2^2+\Omega_3^2+I_1^2+I_2^2+I_3^2\right)E^4
\right.
\nonumber\\
\left.
+\left(
\Omega_1^2\Omega_2^2+\Omega_1^2I_2^2+I_1^2I_2^2
+\Omega_2^2\Omega_3^2+\Omega_2^2I_3^2+I_2^2I_3^2
\right.
\right.
\nonumber\\
\left.
\left.
+\Omega_3^2\Omega_1^2+\Omega_3^2I_1^2+I_3^2I_1^2
\right)E^2
\right.
\nonumber\\
\left.
-\Omega_1^2\Omega_2^2\Omega_3^2
-I_1^2I_2^2I_3^2
\right)^2
\nonumber\\
=-(E^2-a_1)(E^2-a_2)(E^2-a_3)(E^2-a_4)(E^2-a_5)(E^2-a_6),
\label{3.20}
\end{eqnarray}
where  $a_1,\ldots, a_6$
are the solutions of two qubic equations
which follow from the equation
${\cal{A}}_6(E^2)=0$.

Knowing the density of states
$R(E^2)$ (\ref{3.18})
we immediately obtain all quantities of interest.
Thus, the gap in the energy spectrum of the spin model
is given by the square root of the smallest root
of the polynomial ${\cal{A}}_{2p}(E^2)$.
Further,
the ground-state energy per site
is given by
\begin{eqnarray}
e_0=-\int_0^\infty
{\mbox{d}}EE^2R(E^2).
\label{3.21}
\end{eqnarray}
Recalling Eq. (\ref{2.24})
for the Helmholtz free energy per site
we find that
the specific heat per site is given by
\begin{eqnarray}
\frac{c}{k}
=2\int_0^\infty
{\mbox{d}}EER(E^2)
\left(\frac{\frac{\beta E}{2}}{\cosh\frac{\beta E}{2}}\right)^2.
\label{3.22}
\end{eqnarray}
We assume that
$\Omega_n=\Omega+\Delta\Omega_n$
and define
the transverse magnetization per site
and
the static transverse susceptibility per site
as follows:
\begin{eqnarray}
m^z
=\frac{\partial f}{\partial \Omega}
\label{3.23}
\end{eqnarray}
and
\begin{eqnarray}
\chi^z
=\frac{\partial m^z}{\partial \Omega}.
\label{3.24}
\end{eqnarray}

Unfortunately,
the elaborated approach
yields only the density of states (\ref{2.23})
and thus
is restricted to the thermodynamic quantities.
To obtain the spin correlation functions
$\langle s_j^\alpha s_{j+n}^\alpha\rangle$,
$\alpha=x,z$
of the regularly alternating transverse Ising chain
(\ref{2.34}), (\ref{2.35}), (\ref{2.36})
we use the numerical approach.\cite{016,017,024,025}
The on-site magnetizations can be calculated
from the spin correlation functions
using the relation
\begin{eqnarray}
m_{j_1}^\alpha m_{j_2}^\alpha
={\mbox{lim}}_{r\to\infty}
\langle
s^\alpha_{j_1}s^\alpha_{j_2+rp}
\rangle,
\label{3.25}
\end{eqnarray}
where
$1\ll j_1 \ll N$
is one of the $p$ consecutive numbers
taken sufficiently far from the ends of the chain
and
$j_2-j_1=0,1,\ldots,p-1$.
Assuming that
\begin{eqnarray}
\langle
s^\alpha_{j}s^\alpha_{j+rp}
\rangle
-\langle s^\alpha_{j}\rangle
\langle s^\alpha_{j+rp}\rangle
\sim\exp\left(-\frac{rp}{\xi^\alpha}\right)
\label{3.26}
\end{eqnarray}
for large $r$
($1\ll j,j+rp \ll N$)
we can determine the correlation length $\xi^\alpha$.

\section{Effects of Regularly Alternating Bonds/Fields
         on the Quantum Phase Transition
         \label{secd4}}

Now we shall use
the analytical results
for the ground-state and thermodynamic quantities
and the numerical data
for the spin correlation functions
to discuss the effects
of regularly alternating Hamiltonian parameters
on the quantum phase transition
inherent in the transverse Ising chain.

We start with the energy gap $\Delta$.
The quantum phase transition point
is determined by the condition $\Delta=0$.
The density of states $R(E^2)$ permits us
to find the energy gap $\Delta$
since,
as it was mentioned before,
the smallest root of the polynomial ${\cal{A}}_{2p}(E^2)$
(see (\ref{3.18}))
is the smallest elementary excitation energy squared.
Therefore,
the energy spectrum becomes gapless
when
\begin{eqnarray}
{\cal{A}}_{2p}(0)=0.
\label{4.01}
\end{eqnarray}
In the case of period 2
the condition (\ref{4.01}),
${\cal{A}}_{4}(0)=0$,
yields
\begin{eqnarray}
\Omega_1\Omega_2=\pm I_1I_2.
\label{4.02}
\end{eqnarray}
In the case of period 3
the condition (\ref{4.01}),
${\cal{A}}_{6}(0)=0$,
yields
\begin{eqnarray}
\Omega_1\Omega_2\Omega_3=\pm I_1I_2I_3.
\label{4.03}
\end{eqnarray}

It should be emphasized here
that we have rederived the long-known condition
of the zero-energy elementary excitations
obtained by P.~Pfeuty.\cite{031}
P.~Pfeuty showed
that a nonuniform transverse Ising chain becomes gapless if
\begin{eqnarray}
\Omega_1\Omega_2\Omega_3\ldots\Omega_N
=\pm I_1I_2I_3\ldots I_N.
\label{4.04}
\end{eqnarray}
(In fact, Eq. (6) of the P.~Pfeuty's paper\cite{031}
does not contain two signs;
minus immediately follows from symmetry arguments
(after performing simple rotations of spin axes)
and is important for what follows.)

Analysing the conditions for quantum phase transition points
(\ref{4.02}), (\ref{4.03})
we immediately observe
that a number of quantum phase transition points
for a given period of nonuniformity
is strongly conditioned by specific values of the Hamiltonian parameters.
Consider, for example, the case $p=2$
and assume
$\Omega_{1,2}=\Omega\pm\Delta\Omega$,
$\Delta\Omega>0$.
Then,
for a small strength of transverse-field nonuniformity
$\Delta\Omega<\sqrt{\vert I_1I_2\vert}$
the system exhibits two quantum phase transition points
$\Omega_c=\pm\sqrt{\Delta\Omega^2+\vert I_1I_2\vert}$.
Moreover,
the ferromagnetic phase occurs for
$\vert \Omega\vert<\sqrt{\Delta\Omega^2+\vert I_1I_2\vert}$
whereas the paramagnetic phase occurs for
$\sqrt{\Delta\Omega^2+\vert I_1I_2\vert}<\vert \Omega\vert$.
If a strength of transverse-field nonuniformity
is large enough,
$\Delta\Omega>\sqrt{\vert I_1I_2\vert}$,
the system exhibits four quantum phase transition points
$\Omega_c=\pm\sqrt{\Delta\Omega^2\pm\vert I_1I_2\vert}$.
Moreover,
the ferromagnetic phase occurs for
$\sqrt{\Delta\Omega^2-\vert I_1I_2\vert}
<\vert \Omega\vert
<\sqrt{\Delta\Omega^2+\vert I_1I_2\vert}$
whereas the paramagnetic phase occurs
for low fields,
$\vert \Omega\vert<\sqrt{\Delta\Omega^2-\vert I_1I_2\vert}$,
and for strong fields,
$\sqrt{\Delta\Omega^2+\vert I_1I_2\vert}<\vert \Omega\vert$.
In the case
$\Delta\Omega=\sqrt{\vert I_1I_2\vert}$
we get three quantum phase transition points,
$\Omega_c=\pm\sqrt{2}\vert I_1I_2\vert, 0$.
The fields
$\pm\sqrt{2}\vert I_1I_2\vert$
correspond to the transition
between ferromagnetic and paramagnetic phases.
At $\Omega=0$ only a weak singularity occurs.
We shall motivate the statements about the phases
which occur as $\Omega$ varies
considering the dependence
of the ground-state Ising magnetization on $\Omega$
(see below).
The important message
which follows from this simple analysis
is that
the number of quantum phase transitions
in the regularly alternating transverse Ising chain
of a certain period
strongly depends on a specific set of the Hamiltonian parameters.

Let us note
that in our treatment we assume
$\Omega_n=\Omega+\Delta\Omega_n$,
fix $\Delta\Omega_n$ and $I_n$,
and consider the changes in the ground-state properties
as $\Omega$ varies,
thus,
breaking a symmetry between the transverse fields and the exchange interactions.
Alternatively,
we may assume $I_n=I+\Delta I_n$,
fix $\Omega_n$ and $\Delta I_n$,
and assume $I$ to be a free parameter.
In this case the quantum phase transition
is tuned by varying $I$.
Naturally,
in general,
the quantum phase transition conditions (\ref{4.02}), (\ref{4.03})
may be tuned by some parameter(s) effecting
the on-site fields and the intersite interactions.

Let us consider a critical behavior of the system in question.
To get the behavior of the ground-state properties in the vicinity
of the critical fields
we start with expanding the smallest root of the polynomial
${\cal{A}}_{2p}(E^2)$
with respect to deviation of the field from its critical value
$\epsilon=\Omega-\Omega_c$.
For $p=2$ with
$\Omega_{1,2}=\Omega\pm\Delta\Omega$
and $\Omega_c=\pm\sqrt{\Delta\Omega^2\pm\vert I_1I_2\vert}$
we have
\begin{eqnarray}
a_1
=\Omega^2+\Delta\Omega^2
+\frac{1}{2}\left(I_1^2+I_2^2\right)
\nonumber\\
-\sqrt{
\left(\Omega^2+\Delta\Omega^2
+\frac{1}{2}\left(I_1^2+I_2^2\right)\right)^2
-\left(\Omega^2-\Omega_c^2\right)^2
}
\nonumber\\
\approx
\frac{\left(\Omega^2-\Omega_c^2\right)^2}
{2\left(\Omega^2+\Delta\Omega^2
+\frac{1}{2}\left(I_1^2+I_2^2\right)\right)}.
\label{4.05}
\end{eqnarray}
As a result,
the energy gap
$\Delta=\sqrt{a_1}$
decays linearly
\begin{eqnarray}
\Delta\sim\vert\epsilon\vert
\label{4.06}
\end{eqnarray}
if $\Omega_c\ne 0$
and as
\begin{eqnarray}
\Delta\sim\epsilon^2
\label{4.07}
\end{eqnarray}
if $\Omega_c=0$.
The latter case occurs
when $\Delta\Omega=\sqrt{\vert I_1I_2\vert}$.

\begin{figure}[th]
\centerline{\psfig{file=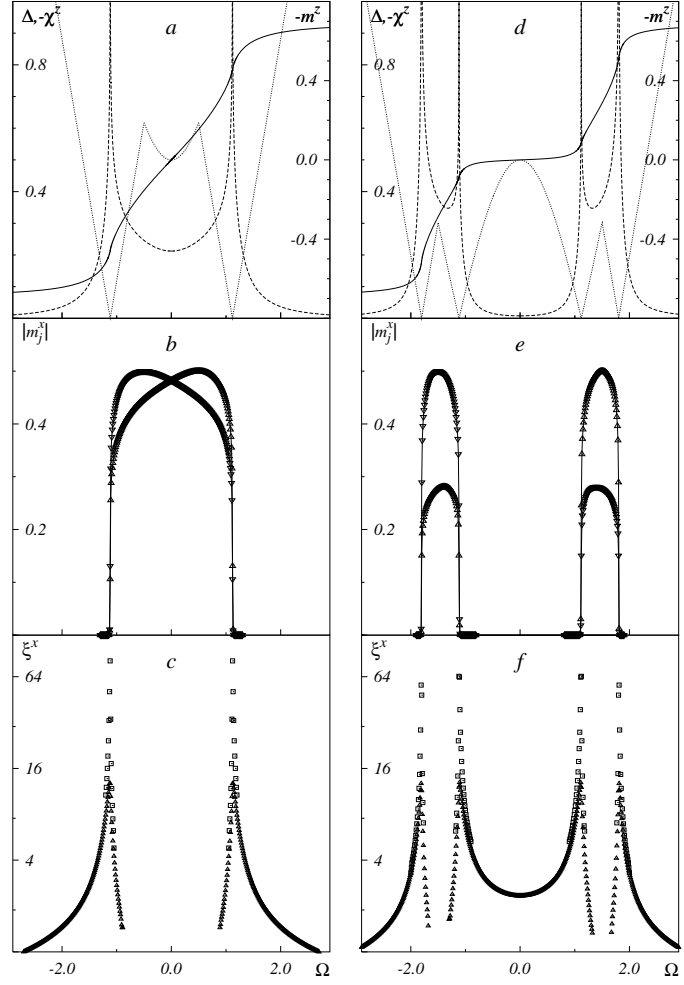,width=3.65in,angle=0}} 
\vspace*{8pt}
\caption{
Towards the ground-state properties
of the transverse Ising chain of period 2,
$I_1=I_2=1$,
$\Omega_{1,2}=\Omega\pm\Delta\Omega$,
$\Delta\Omega=0.5$
($\Omega_c=\pm\sqrt{1.25}$)
(a - c),
$\Delta\Omega=1.5$
($\Omega_c=\pm\sqrt{1.25},\pm\sqrt{3.25}$)
(d - f).
The dependences of
energy gap
(dotted curves in (a), (d)),
transverse magnetization $m^z$
(solid curves in (a), (d)),
static transverse susceptibility $\chi^z$
(dashed curves in (a), (d)),
longitudinal sublattice magnetizations $\vert m_j^x\vert$
(solid curves with up and down triangles in (b), (e)
which are obtained from the data for $N=600$),
and
correlation length $\xi^x$
(triangles and squares in (c), (f)
correspond to the data for $N=300$ and $N=600$, respectively),
on the transverse field $\Omega$.}
\label{fig04}
\end{figure}

Now we can evaluate the ground-state energy (\ref{3.21})
in the vicinity of the critical point.
The nonanalytic contribution
in the r.h.s. of Eq. (\ref{3.21})
is coming from the energy interval
between $\sqrt{a_1}$
(which is proportional either to $\vert\epsilon\vert$
or to $\epsilon^2$)
and $\sqrt{a_2}$ ($a_1<a_2<a_3<\ldots$);
all the other intervals of energies yield only analytical contributions
to the ground-state energy $e_0$ with respect to $\epsilon$.
Therefore,
we have
\begin{eqnarray}
e_0=
-\frac{1}{2\pi}
\int_{\sqrt{a_1}}^{\sqrt{a_2}}
{\mbox{d}}E\frac{E^2f(E^2)}{\sqrt{E^2-a_1}}
+{\mbox{analytical with respect to $\epsilon^2$ terms}}
\nonumber\\
\sim\epsilon^2\ln\vert \epsilon \vert
+{\mbox{analytical with respect to $\epsilon^2$ terms}}.
\nonumber\\
\label{4.08}
\end{eqnarray}
(Here $f(E^2)$, $f(0)\ne 0$
is some function
the explicit expression of which is not important
for deriving the asymptotic behavior.)
Eq. (\ref{4.08}) is valid for $\Omega_c\ne 0$
when $\sqrt{a_1}\sim\vert\epsilon\vert$.
As a result
\begin{eqnarray}
m^z
\sim\epsilon\ln\vert\epsilon\vert
+{\mbox{analytical with respect to $\epsilon^2$ terms}},
\label{4.09}
\end{eqnarray}
and
\begin{eqnarray}
\chi^z
\sim\ln\vert\epsilon\vert
+{\mbox{analytical with respect to $\epsilon^2$ terms}}.
\label{4.10}
\end{eqnarray}
Thus,
the critical behavior remains unchanged
in comparison with that of the uniform chain
(compare with (\ref{2.42}), (\ref{2.43}), (\ref{2.44}), (\ref{2.45})).
This can be also nicely seen in Fig.~\ref{fig04}
where some results
referring to the chains of period 2
with
$I_1=I_2=1$
and
$\Delta\Omega=0.5$ and $\Delta\Omega=1.5$
are collected.

Let us turn now to the case $\Omega_c=0$
which occurs for the chain of period 2
with $\Delta\Omega=\sqrt{\vert I_1I_2\vert}$.
In this case the smallest root $a_1$
tends to zero proportionally to $\epsilon^4$
as $\Omega$ approaches $\Omega_c$
and as a result
\begin{eqnarray}
e_0
\sim\epsilon^4\ln\vert\epsilon\vert
+{\mbox{analytical terms}}
\label{4.11}
\end{eqnarray}
and, therefore,
\begin{eqnarray}
m^z
\sim\epsilon^3\ln\vert\epsilon\vert
+{\mbox{analytical terms}}
\label{4.12}
\end{eqnarray}
and
\begin{eqnarray}
\chi^z
\sim\epsilon^2\ln\vert\epsilon\vert
+{\mbox{analytical terms}}.
\label{4.13}
\end{eqnarray}
Thus,
the transverse susceptibility $\chi^z$ is finite
in the critical point $\Omega_c=0$
and only its second derivative with respect to the field
$\frac{\partial^2\chi^z}{\partial\Omega^2}$
exhibits
a logarithmic singularity.
This weak singularity may be called
the fourth-order quantum phase transition
in the Ehrenfest sense.

I proceed with exact analytical results
for a regularly alternating transverse Ising chain
showing how the ground-state wave function
$\vert {\mbox{GS}}\rangle$
can be obtained exactly
for special sets of the Hamiltonian parameters.
Let us consider the chain of period 2 with
$\Omega_{1,2}=\Omega\pm\Delta\Omega$.
It is obvious that
\begin{eqnarray}
\vert {\mbox{GS}}\rangle
=\ldots \vert\uparrow\rangle_j\vert\uparrow\rangle_{j+1}\ldots
\label{4.14}
\end{eqnarray}
as $\Omega\to-\infty$
and
\begin{eqnarray}
\vert {\mbox{GS}}\rangle
=\ldots \vert\downarrow\rangle_j\vert\downarrow\rangle_{j+1}\ldots
\label{4.15}
\end{eqnarray}
as $\Omega\to\infty$.
We can also expect that at $\Omega=0$
\begin{eqnarray}
\vert {\mbox{GS}}\rangle
=\ldots \vert\downarrow\rangle_j\vert \uparrow\rangle_{j+1}\ldots
\label{4.16}
\end{eqnarray}
if $\Delta\Omega\gg \vert I_1\vert,\vert I_2\vert$
and
\begin{eqnarray}
\vert {\mbox{GS}}\rangle
=\ldots
\frac{1}{\sqrt{2}}
\left(\vert\downarrow\rangle_j\pm\vert\uparrow\rangle_j\right)
\frac{1}{\sqrt{2}}
\left(\vert\downarrow\rangle_{j+1}\pm\vert\uparrow\rangle_{j+1}\right)
\ldots
\label{4.17}
\end{eqnarray}
if $\Delta\Omega=0$.
Less evident
is the ground-state wave function $\vert {\mbox{GS}}\rangle$ for
$\Omega=\mp\Delta\Omega$
when
$\Omega_1=0$, $\Omega_2=-2\Delta\Omega$
or
$\Omega_1=2\Delta\Omega$, $\Omega_2=0$.
To obtain the ground-state wave functions
for these values of the transverse field
let us note
that after performing
the unitary transformations
\begin{eqnarray}
U=\ldots
\exp\left({\mbox{i}}\pi s_n^x s_{n+1}^y\right)
\exp\left({\mbox{i}}\pi s_{n+1}^x s_{n+2}^y\right)
\ldots
\label{4.18}
\end{eqnarray}
and
\begin{eqnarray}
R^z=\ldots
\exp\left({\mbox{i}}\frac{\pi}{2} s_n^z\right)
\exp\left({\mbox{i}}\frac{\pi}{2} s_{n+1}^z\right)
\ldots
\label{4.19}
\end{eqnarray}
the Hamiltonian of the transverse Ising chains
with the fields
$\Omega_1\Omega_2\ldots$
and the interactions
$I_1I_2\ldots$
transforms
(with the accuracy to the boundary terms)
into
the Hamiltonian of the transverse Ising chains
with the fields
$I_1I_2\ldots$
and the interactions
$\Omega_1\Omega_2\ldots\;$,
\begin{eqnarray}
R^zUH{U}^+{R^z}^+
=
\sum_jI_js^z_j+\sum_j2\Omega_{j+1}s_j^xs_{j+1}^x.
\label{4.20}
\end{eqnarray}
For
$\Omega=\mp\Delta\Omega$
the Hamiltonian
$R^zUH{U}^+{R^z}^+$
represents a system of noninteracting clusters
which consist of two sites.
We can easily find the ground-state
of a two-site cluster
with the Hamiltonian
($\Omega=-\Delta\Omega$)
\begin{eqnarray}
H_{12}=I_1s_1^z+I_2s_2^z
-4\Delta\Omega s_1^xs_2^x.
\label{4.21}
\end{eqnarray}
It reads ($I_1I_2>0$)
\begin{eqnarray}
\vert {\mbox{GS}}\rangle_{12}
=c_1\vert\downarrow\rangle_1\vert\downarrow\rangle_2
+
c_2\vert\uparrow\rangle_1\vert\uparrow\rangle_2,
\nonumber\\
c_1=\frac{1}{\sqrt{2}}
\frac{I+\sqrt{I^2+\Delta\Omega^2}}{\sqrt{I^2+I\sqrt{I^2+\Delta\Omega^2}+\Delta\Omega^2}},
\nonumber\\
c_2=\frac{1}{\sqrt{2}}
\frac{\Delta\Omega}{\sqrt{I^2+I\sqrt{I^2+\Delta\Omega^2}+\Delta\Omega^2}},
\nonumber\\
I=\frac{1}{2}\left(I_1+I_2\right).
\label{4.22}
\end{eqnarray}
As a result,
the desired ground-state wave function reads
\begin{eqnarray}
\vert {\mbox{GS}}\rangle
=U^+{R^z}^+\ldots
\left(
c_1\vert\downarrow\rangle_{n+1}\vert\downarrow\rangle_{n+2}
+
c_2\vert\uparrow\rangle_{n+1}\vert\uparrow\rangle_{n+2}
\right)
\nonumber\\
\cdot
\left(
c_1\vert\downarrow\rangle_{n+3}\vert\downarrow\rangle_{n+4}
+
c_2\vert\uparrow\rangle_{n+3}\vert\uparrow\rangle_{n+4}
\right)
\ldots .
\label{4.23}
\end{eqnarray}
For $\Omega=\Delta\Omega$,
the ground-state is again given by (\ref{4.23}), (\ref{4.22})
with the change
$n\to n-1$,
$\Delta\Omega\to -\Delta\Omega$.

Knowing the ground-state wave function
we can easily calculate different spin correlation functions.
For example,
for
$\Omega=-\Delta\Omega$
according to Eq. (\ref{4.23}) we get
\begin{eqnarray}
\langle s_{n+1}^x s_{n+2}^x \rangle
=\langle s_{n+1}^x s_{n+4}^x \rangle
=\ldots
\nonumber\\
=\langle s_{n+2}^x s_{n+3}^x \rangle
=\langle s_{n+2}^x s_{n+5}^x \rangle
=\ldots
\nonumber\\
=\frac{1}{4}\left(c_2^2-c_1^2\right),
\nonumber\\
\langle s_{n+1}^x s_{n+3}^x \rangle
=\langle s_{n+1}^x s_{n+5}^x \rangle
=\ldots
\nonumber\\
=\frac{1}{4}\left(c_2^2-c_1^2\right)^2,
\nonumber\\
\langle s_{n+2}^x s_{n+4}^x \rangle
=\langle s_{n+2}^x s_{n+6}^x \rangle
=\ldots
=\frac{1}{4};
\label{4.24}
\end{eqnarray}
\begin{eqnarray}
\langle s_{n+1}^z s_{n+2}^z \rangle
=\langle s_{n+1}^z s_{n+4}^z \rangle=\ldots
\nonumber\\
=\langle s_{n+2}^z s_{n+3}^z \rangle
=\langle s_{n+2}^z s_{n+5}^z \rangle
=\ldots
\nonumber\\
=\langle s_{n+2}^z s_{n+4}^z \rangle
=\langle s_{n+2}^z s_{n+6}^z \rangle
=\ldots
=0,
\nonumber\\
\langle s_{n+1}^z s_{n+3}^z \rangle
=\langle s_{n+1}^z s_{n+5}^z \rangle
=\ldots
=\left(c_1c_2\right)^2.
\label{4.25}
\end{eqnarray}
For $\Omega=\Delta\Omega$,
the correlation functions follow from
(\ref{4.24}), (\ref{4.25})
after the change $n\to n-1$.

\begin{figure}[th]
\centerline{\psfig{file=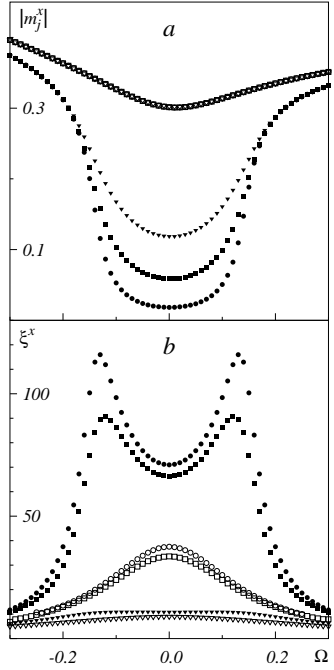,width=2.0in,angle=0}}
\vspace*{8pt}
\caption{Towards the ground-state properties
of the transverse Ising chain of period 2,
$I_1=I_2=1$,
$\Omega_{1,2}=\Omega\pm\Delta\Omega$,
$\Delta\Omega=0.99$ (open symbols),
$\Delta\Omega=1.01$ (full symbols).
The longitudinal sublattice magnetization $\vert m_j^x\vert$ (a)
and
the correlation length $\xi^x$ (b)
against
the transverse field $\Omega$
for chains having 300, 600 and 900 sites
(triangles, squares and circles).}
\label{fig05}
\end{figure}

Now we discuss the exact numerical results for finite chains
presented in Figs.~\ref{fig04},~\ref{fig05}.
The ground-state Ising (longitudinal) sublattice magnetizations
$\vert m^x_j\vert$
indicate different phases and phase transitions between them.
Their dependence on the transverse field $\Omega$
are  shown in Figs.~\ref{fig04},~\ref{fig05}
for
$\Delta\Omega<\sqrt{\vert I_1I_2\vert}$
(Fig.~\ref{fig04}b and Fig.~\ref{fig05}a (open symbols))
and
$\Delta\Omega>\sqrt{\vert I_1I_2\vert}$
(Fig.~\ref{fig04}e and Fig.~\ref{fig05}a (full symbols)).
In accordance with the analytical results for $\Delta$, $m^z$
and $\chi^z$
the numerical data show the existence of
either two phases
(quantum Ising (ferromagnetic) phase for
$\vert \Omega\vert <\sqrt{\Delta\Omega^2+\vert I_1 I_2\vert}$
and strong-field quantum paramagnetic phase otherwise)
or three phases
(low-field quantum paramagnetic phase for
$\vert \Omega\vert <\sqrt{\Delta\Omega^2-\vert I_1 I_2\vert}$,
quantum Ising (ferromagnetic) phase for
$\sqrt{\Delta\Omega^2-\vert I_1 I_2\vert}
<\vert \Omega\vert
<\sqrt{\Delta\Omega^2+\vert I_1 I_2\vert}$
and strong-field quantum paramagnetic phase otherwise).
The longitudinal on-site magnetizations $m_j^x$
are nonzero
in quantum Ising phases
and become strictly zero
in quantum paramagnetic phases.
The transverse magnetization $m^z$ in quantum paramagnetic phases
is almost constant
being in the vicinity
of zero in the low-field phase
and
of saturation value in the strong-field phase,
thus producing plateau-like steps in the dependence $m^z$ versus $\Omega$
(see the solid curve in Fig.~\ref{fig04}d).
The  correlation length $\xi^x$
(Figs.~\ref{fig04}c,~\ref{fig04}f,~\ref{fig05}b (full symbols))
illustrates that the transition between different phases
is accompanied by the divergency of $\xi^x$.
Comparing the data for $N=600$ and $N=900$
reported in Fig.~\ref{fig05}b
one observes
a strong size-dependence of the correlation length value $\xi^x$
about the critical fields
(full squares and circles in Fig.~\ref{fig05}b)
and a weak size-dependence of $\xi^x$
for other values of $\Omega$
(for example, open squares and circles in Fig.~\ref{fig05}b).
It is interesting to note
that
short chains ($N=20$) are already sufficient to
reproduce a correct order-parameter behavior
in the quantum Ising phase away from the quantum critical point,
but not in the quantum paramagnetic phases.\cite{024}
The chain length of up to
$N=900$ sites is clearly sufficient to see a sharp transition in the order
parameter (Figs.~\ref{fig04}b,~\ref{fig04}e)
and even to extract reliable results for $\xi^x$
from the long-distance  behavior of $\langle s_j^{x} s_{j+n}^{x} \rangle$
as can be seen from the data  presented
in Figs.~\ref{fig04}c,~\ref{fig04}f
and in Fig.~\ref{fig05}b.
We also note that the numerical data for
$\vert m_j^x\vert$ and $\xi^x$ at $\Omega=\mp\Delta\Omega$
coincide with the exact expression (\ref{4.24}).
In particular,
the values of the longitudinal sublattice magnetizations
for these fields
are
$\frac{1}{2}$
and
$\frac{1}{2}\vert c_2^2-c_1^2\vert$.

\begin{figure}[th]
\centerline{\psfig{file=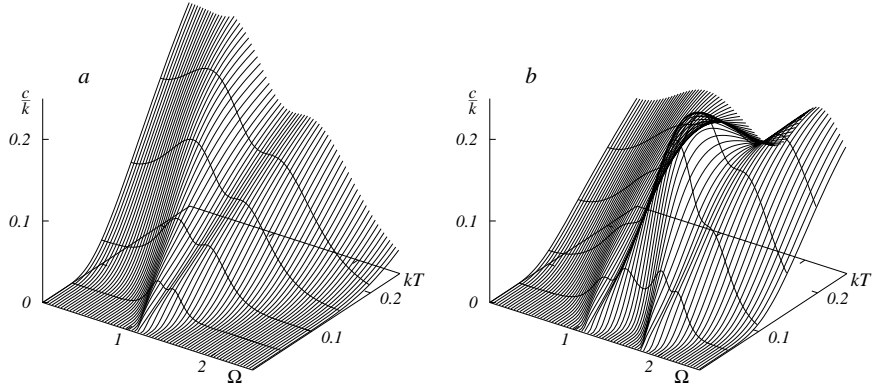,width=4.7in,angle=0}}
\vspace*{8pt}
\caption{The low-temperature dependence
of the specific heat
for the transverse Ising chain of period 2,
$I_1=I_2=1$,
$\Omega_{1,2}=\Omega\pm\Delta\Omega$,
$\Delta\Omega=0.5$ (a),
$\Delta\Omega=1.5$ (b).}
\label{fig06}
\end{figure}

Finally,
the low-temperature dependence of the specific heat $c$
at different $\Omega$ ($\ge 0$)
obtained from the exact analytical expression (\ref{3.22})
confirms the
existence of
either two phase transitions
(Fig.~\ref{fig06}a)
or four phase transitions
(Fig.~\ref{fig06}b)
depending on the relationship between
$\Delta\Omega$ and $\sqrt{\vert I_1I_2\vert}$.
At the quantum phase transition points
the spin chain becomes gapless
and $c$ depends linearly on $T$.
Really,
at the quantum phase transition point
as $T\to 0$
\begin{eqnarray}
\frac{c}{k}
=
2\int_0^{kT}{\mbox{d}}E E\frac{C}{\sqrt{E^2}}
\left(\frac{E}{2kT}\right)^2
+
2\int_{kT}^\infty{\mbox{d}}E ER(E^2)
\left(\frac{\frac{E}{2kT}}{\cosh\frac{E}{2kT}}\right)^2.
\nonumber\\
\label{4.26}
\end{eqnarray}
(Here $C$ is some constant
the value of which is not important
for deriving the asymptotic behavior.)
The second term in (\ref{4.26})
disappears in the limit $T\to 0$
($E>kT$)
and as a result
\begin{eqnarray}
c\sim T.
\label{4.27}
\end{eqnarray}

The ridges seen in Figs.~\ref{fig06}a,~\ref{fig06}b
correspond to the maxima in the dependence $c$ versus $\Omega$
as $T$ varies.
They single out the boundaries
of quantum critical regions.\cite{008}
These boundaries correspond to a relation  $\Delta\sim kT$
that can be checked
by comparison with the data for $\Delta$ versus $\Omega$
reported in Figs.~\ref{fig04}a,~\ref{fig04}d.
As can be seen from Fig.~\ref{fig06}
the $c(T)$ behavior for $\Omega$
slightly above or below $\Omega_c$
changes crossing the boundaries of quantum critical regions.
Furthermore,
we notice that for $\Delta\Omega=1.5$
(Fig.~\ref{fig06}b)
an additional low-temperature peak appears
in the temperature dependence of the specific heat.

\begin{figure}[th]
\centerline{\psfig{file=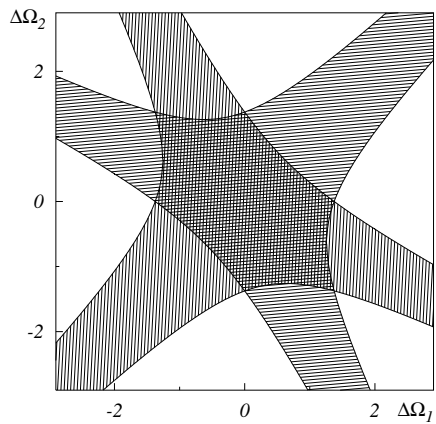,width=2.5in,angle=0}}
\vspace*{8pt}
\caption{Towards the number of quantum phase transitions
present in the transverse Ising chain of period 3,
$I_1=I_2=I_3=1$,
$\Omega_{1,2,3}=\Omega+\Delta\Omega_{1,2,3}$,
$\Delta\Omega_1+\Delta\Omega_2+\Delta\Omega_3=0$.
The dark, gray, or light regions correspond to parameters
for which the system exhibits two, four, or six quantum phase transitions,
respectively.
At the boundaries between different regions
weak singularities occur.}
\label{fig07}
\end{figure}

To end up,
let us turn to a chain of period 3
for which the analytical and the numerical calculations
presented above can be repeated.
The quantum phase transition points follow from the condition
(\ref{4.03}).
Depending on the parameters,
either two, four, or six
quantum phase transitions are possible.
This behavior is illustrated in Fig.~\ref{fig07}
for a chain of period 3
with
$I_1=I_2=I_3=1$,
$\Omega_{n}=\Omega+\Delta\Omega_n$,
$\Delta\Omega_1+\Delta\Omega_2+\Delta\Omega_3=0$.
Moreover,
such chains may exhibit weak singularities.
For example,
for the set of parameters at the boundary between dark and gray regions
there are three critical fields;
at one of them a weak singularity occurs.

To summarize,
I have shown how the exact results
for the thermodynamics
of a regularly alternating
spin-$\frac{1}{2}$ Ising chain in a transverse field
can be derived using the Jordan-Wigner transformation
and continued fractions.
Furthermore,
for special parameter values
the exact ground-state wave function and spin correlation functions
can be obtained as well.
For parameters
for which the correlation functions are not accessible
for rigorous analytical analysis
they can be obtained from
exact numerical results
for finite but large systems.
From these numerical data
we can extract the order parameter and the correlation length
in high precision.
We have found,
that the quantum phase transition occurring
in the uniform transverse Ising chain
is not suppressed by deviation from uniformity
in the form of a regular alternation of bonds and fields.
On the contrary,
field alternation may lead
to the appearance of additional quantum phase transitions
tuned by the field.
The number of phase transitions
for a given period of alternation
strongly depends
on the precise values of the parameters of the model.
We have examined the critical behavior
of the energy gap $\Delta$,
the ground-state energy $e_0$,
the transverse magnetization $m^z$,
and
the static transverse susceptibility $\chi^z$
and have found the same critical indices
both
for a nonuniform and uniform chain.
For some sets of the Hamiltonian parameters,
the ground-state quantities
may exhibit a weak singularity.

\section{Related Models
         \label{secd5}}

In the remainder of this chapter
I shall discuss another simple models
which,
apart from the transverse Ising chain,
exhibit a quantum phase transition.
Namely,
I consider
the spin-$\frac{1}{2}$ isotropic $XY$
(i.e. $XX$)
chain in a transverse field
and
the spin-$\frac{1}{2}$ $XYZ$ chain in an external magnetic field
directed along $z$ axis
focusing on
the effects of regularly alternating Hamiltonian parameters
on the dependence of the ground-state quantities
on the external field.

The uniform transverse $XX$ chain
described by the Hamiltonian
\begin{eqnarray}
H
=\sum_{n=1}^N\Omega s_n^z
+\sum_{n=1}^N2I\left(s_n^xs_{n+1}^x+s_n^ys_{n+1}^y\right)
\label{5.01}
\end{eqnarray}
by employing the Jordan-Wigner transformation
(\ref{2.06}), (\ref{2.07})
can be mapped onto the system of noninteracting spinless fermions
with the Hamiltonian
\begin{eqnarray}
H
=\sum_{\kappa=1}^N
\Lambda_\kappa
\left(c^+_\kappa c_\kappa-\frac{1}{2}\right)
\label{5.02}
\end{eqnarray}
where the elementary excitation energies
$\Lambda_\kappa$
are given by
\begin{eqnarray}
\Lambda_\kappa=\Omega+2I\cos\kappa.
\label{5.03}
\end{eqnarray}
Alternatively,
we may find the density of states $\rho(E)$ (\ref{2.22})
which is given by
\begin{eqnarray}
\rho(E)
=\left\{
\begin{array}{ll}
\frac{1}{\pi\sqrt{4I^2-\left(E-\Omega\right)^2}}, &
\;\;\;
{\mbox{if $4I^2-\left(E-\Omega\right)^2>0$,}} \\
0, &
\;\;\;
{\mbox{otherwise.}}
\end{array}
\right.
\label{5.04}
\end{eqnarray}
As a result,
the thermodynamic properties of the spin chain (\ref{5.01})
can be easily analyzed.

In contrast to the transverse Ising chain
(and to any chain with anisotropic $XY$ interaction in a transverse field)
$\sum_{n=1}^Ns_n^z$
commutes with the Hamiltonian
describing the isotropic $XY$ interspin interaction
and, therefore,
a change in the transverse field $\Omega$
has a different effect
on the ground-state quantities of the spin chain (\ref{5.01}).
As can be seen from (\ref{5.02}), (\ref{5.03})
the transverse field plays a role of the chemical potential
controlling a filling of the fermion band.
The transverse $XX$ chain remains gapless until
$\vert\Omega\vert\le 2\vert I\vert$.
If $\vert\Omega\vert$ exceeds $2\vert I\vert$,
the energy gap $\Delta$ opens linearly.
This produces singularities
in the ground-state quantities.
For example,
a square-root singularity
in the zero-temperature dependence
of the static transverse susceptibility $\chi^z$
on the transverse field $\Omega$.

After introducing a regular alternation,
i.e. after making a substitution in (\ref{5.01})
$\Omega\to\Omega_n$,
$I\to I_n$
with a periodic sequence of parameters
\begin{eqnarray}
\Omega_1I_1\Omega_2I_2\ldots\Omega_pI_p
\Omega_1I_1\Omega_2I_2\ldots\Omega_pI_p\ldots ,
\nonumber
\end{eqnarray}
we can easily obtain with the help of continued fractions
the density of states $\rho(E)$
of the Jordan-Wigner fermions
which represent the regularly alternating transverse $XX$ chain,
and hence
analyze the effects of regularly alternating Hamiltonian parameters
on the thermodynamic properties of the considered spin system\cite{032}
(for another approach see Ref. \refcite{033}).
The main consequence of the introduced periodic nonuniformity
is a splitting of the initial fermion band
into several subbands
the number of which does not exceed the period of the chain $p$
(for special (symmetric) values of the Hamiltonian parameters
one may observe
a smaller than $p$ number of subbands).
The transverse field $\Omega$
($\Omega_n=\Omega+\Delta\Omega_n$)
controls a filling of the fermion subbands.
Again the energy gap $\Delta$ disappears/appears linearly
and the critical behavior remains
as for the uniform chain.
Note,
that the ground-state dependence of the transverse magnetization
\begin{eqnarray}
m^z=-\frac{1}{2}
\int_{-\infty}^{\infty}{\mbox{d}}E\rho(E)\tanh\frac{E}{2kT}
\label{5.05}
\end{eqnarray}
on $\Omega$
is composed of sharply increasing parts
separated by horizontal parts (plateaus)
in accordance with a famous conjecture
of M.~Oshikawa, M.~Yamanaka and I.~Affleck
(see Ref. \refcite{034}).

A more general situation
is the case of regularly alternating $XYZ$ chain in a magnetic field.
Let us consider the Hamiltonian
\begin{eqnarray}
H
=\sum_n\Omega s_n^z
+\sum_n
\left(
I_n^xs_n^xs_{n+1}^x+I_n^ys_n^ys_{n+1}^y
+I^z s_n^zs_{n+1}^z
\right),
\nonumber\\
I_n^x=1+\gamma +\left(-1\right)^n\delta,
\;\;\;
I_n^y=1-\gamma +\left(-1\right)^n\delta,
\label{5.06}
\end{eqnarray}
which captures the interplay between
an exchange interaction anisotropy ($\gamma$, $I^z$),
exchange interaction modulation
or, more precisely, exchange interaction dimerization ($\delta$),
and the uniform magnetic field ($\Omega$).
Such a model has been studied recently
in some details\cite{035}
using the bosonization approach\cite{036,037,038}
and the Lanczos diagonalization technique.
Let us first discuss the free fermion case $I^z=0$.
After the Jordan-Wigner transformation
(\ref{2.06}), (\ref{2.07})
the Hamiltonian can be readily diagonalized
(see Refs. \refcite{026,027,039,035}).
The critical field values
at which the spin system (\ref{5.06}) becomes gapless
are given by
\begin{eqnarray}
\Omega_c=\pm\sqrt{\delta^2-\gamma^2}.
\label{5.07}
\end{eqnarray}
At the critical field
the (transverse) magnetization behaves as
\begin{eqnarray}
m^z-m^z_c
\sim
\left(\Omega-\Omega_c\right)
\left(\ln\vert\Omega-\Omega_c\vert-1\right)
\label{5.08}
\end{eqnarray}
and the static (transverse) susceptibility $\chi^z$
exhibits a logarithmic singularity
(compare with Eqs. (\ref{4.09}), (\ref{4.10})).
The bosonization approach results\cite{035}
show that the same picture is valid for an arbitrary $I^z\ne 0$
provided $\delta$ and $\gamma$ are suitably renormalized.
The results of bosonization approach
elaborated for small $\gamma$ and $\delta$
may be compared with numerical findings
within non-perturbative regimes.\cite{035}
From this analysis
($0\le I^z\le 1$)
it was found
that the critical line for $\Omega=0$ is given by
\begin{eqnarray}
\delta\sim\gamma
\label{5.09}
\end{eqnarray}
in accord with (\ref{5.07}).
For the behavior of the magnetization curves
near the critical fields $\Omega_c$
($I^z=1$)
a fair regime (\ref{5.08}) was obtained.
These outcomes suggest
that some basic features of the fully interacting system (\ref{5.06})
are captured by the free fermion picture.

\section*{Acknowledgments}
\addcontentsline{toc}{section}{Acknowledgments}

I would like to thank
Johannes~Richter,
Taras~Krokhmalskii
and
Oles'~Zaburannyi
in collaboration with whom
the study of the regularly alternating transverse Ising chains
was performed.
I also would like to thank
the DFG for the support of my work
over the past five years in a number of ways.
Finally,
I thank
doctor Janush~Sanotsky
owing to whom
I was able to give a talk at the Ising Lectures - 2002.

\end{document}